\documentclass[aps,prb,showpacs,floatfix,twocolumn]{revtex4-1}
\usepackage{amssymb}
\usepackage{graphicx}
\DeclareGraphicsExtensions{.pdf}
\usepackage{color}
\usepackage{amsmath}
\usepackage{amsbsy}
\usepackage{url}
\usepackage[colorlinks=true]{hyperref}
\hypersetup{
     linktocpage=true,
     hyperindex=true,
     bookmarks=true,
     bookmarksopen=true,
     bookmarksnumbered=true,
     linkcolor=blue,
     anchorcolor=blue,
     citecolor=blue,
     filecolor=blue,
     menucolor=blue,
     runcolor=blue,
     urlcolor=blue,
     pdfborder=0 0 0,
}

\begin{document}

\date{October 27, 2011}

\title{Interaction of Josephson and magnetic oscillations in Josephson\\
tunnel junctions with a ferromagnetic layer}

\author{S.~Mai}
\author{E.~Kandelaki}
\author{A.F.~Volkov}
\author{K.B.~Efetov}
\affiliation{Theoretische Physik III, Ruhr-Universit\"{a}t Bochum, D-44780 Bochum, Germany}

\begin{abstract}
We study the dynamics of Josephson junctions with a thin ferromagnetic layer F
[superconductor-ferromagnet-insulator-ferromagnet-superconductor (SFIFS) junctions].
In such junctions, the phase difference $\varphi$ of the superconductors and magnetization
$M$ in the F layer are two dynamic parameters coupled to each other. We
derive equations describing the dynamics of these two parameters and formulate the
conditions of validity. The coupled Josephson
plasma waves and oscillations of the magnetization $M$ affect the form of the
current-voltage ($I$-$V$) characteristics in the presence of a weak magnetic field (Fiske steps).
We calculate the modified Fiske steps and show that the magnetic degree of
freedom not only changes the form of the Fiske steps but also the
overall view of the $I$-$V$ curve (new peaks related to the magnetic resonance
appear). The $I$-$V$ characteristics are shown for different lengths of the
junction including those which correspond to the current experimental
situation. We also calculate the power $P$ absorbed in the system if a microwave
radiation with an ac in-plane magnetic field is applied (magnetic
resonance). The derived formula for the power $P$ essentially differs from
the one which describes the power absorption in an isolated ferromagnetic
film. In particular, this formula describes the peaks related to the
excitation of standing plasma waves as well as the peak associated with the
magnetic resonance.
\end{abstract}

\maketitle

\section{Introduction}
\label{section-Intro}

A great attention in recent years has been paid to the study of Josephson
junctions (JJ) with a magnetic layer (or layers) \cite{GolubovRMP,BuzdinRMP,BVErmp,PhysToday}.
Although the exchange field in the ferromagnetic layer F essentially suppresses
the Josephson current $I_{J}$,
the interaction of the exchange field and singlet Cooper pairs results in
new, interesting, and nontrivial effects. For example, the singlet pair wave function
penetrating from the superconducting leads into the F layer due to the proximity effect
oscillates in space. In case of a uniform F layer, the pair wave function consists of two
components: one is the singlet component and another is the triplet component
with zero projection of the total spin on the direction of the magnetization
vector $\mathbf{M}$ in the ferromagnet. The condensate wave function decays
in the ferromagnet on a short distance from the superconductor-ferromagnet (SF) interfaces, which in the
diffusive limit, is of the order $\xi _{F}=\sqrt{D/2E_{\mathrm{exc}}}$, where $D=v_Fl/3$
is the diffusion constant and $E_{\mathrm{exc}}$ is the exchange energy.
Here, $v_F$ and $l$ denote the Fermi velocity and the electron mean free path, respectively.
Oscillations of the Cooper pair wave function in space lead to a change of sign
of the critical Josephson current $I_{Jc}$. This effect was predicted long
ago \cite{Bulaev,BuzdinSFS} but observed only recently
\cite{Ryaz01,Ryaz06,Kontos,Blum,Bauer,Sellier,PalVolkovEfetov,Weides09}.

If the magnetization in the F layer is not uniform (for example, this occurs
in the case of a domain structure or multilayered ferromagnet-superconductor (FS) structures with
noncollinear magnetization directions in the F layers), due to the proximity effect
a so-called odd-frequency triplet component arises \cite{BVErmp,PhysToday,EschrigR}.
In contrast to a conventional triplet component that is
an odd function of momentum and is suppressed by scattering off ordinary impurities
\cite{Maeno}, the odd-frequency triplet component is an even function of momentum (in the
diffusive case) and is not destroyed by scattering off ordinary impurities.
This component also is not sensitive to the exchange field and therefore can
penetrate into the ferromagnet over a long distance up to $\xi_{N}=
\sqrt{D/2\pi T}$ at temperature $T$. Convincing data in favor of existence of this long-range
triplet component have been obtained in a number of recent experimental
works \cite{Keizer,Sosnin,Birge,Chan,BlamireScience,Westerholt10,Aarts}.

Another interesting effect arises in SFIFS junctions. It turns out that at
the antiferromagnetic magnetization orientation in the F layers, the
Josephson critical current $I_{Jc}$ is increased \cite{BVEcrcur}. Its value
may even exceed the critical current $I_{Jc}$ in similar JJs without
ferromagnetic layers. This prediction was also confirmed experimentally
\cite{BlamirePRL}.

Alongside with the study of the dc Josephson current in SFS or S$\mathrm{F}_1\mathrm{F}_2\mathrm{F}_1$S JJs,
dynamic properties of these junctions and also of tunnel SIFS or SFIFS JJs
have been investigated both experimentally \cite{Weides06,Ustinov,Wild,Weides10} and theoretically \cite{VE,Maekawa,Aprili}.
Here and throughout the paper, S and I, respectively, represent a superconducting and insulating layers and
$\mathrm{F}_{1/2}$ denotes two distinct ferromagnetic layers.
Interesting dynamic phenomena in JJs with a ferromagnetic layer or a magnetic
particle occur when the dynamics of the superconducting phase difference $%
\varphi(t)$ and the magnetization $M(t)$ come into play.

The coupling between these two degrees of freedom may be realized in
different ways. For example, the Josephson current produces a torque acting
on magnetization vectors in multilayered S$\mathrm{F}_1\mathrm{F}_2$S junctions. Since the
Josephson current $I_{J}[\varphi (t)]$ is determined by the mutual
orientations of magnetization vectors $\mathbf{M}_{1/2}$, the dynamic behavior of the Josephson current will depend on
the dynamics of $M(t)$\cite{Waintal,Braude,Sauls}. Another mechanism of the
supercurrent action on magnetization was considered by Konschelle and Buzdin %
\cite{Konschelle}. They studied dynamics of SFS junctions with a
non-centrosymmetric ferromagnet. In this case, the Josephson current $I_{J}$
acts directly on the magnetization $M$ leading to its precession. In a
nonstationary case, the interplay between $I_{J}(t)$ and $M(t)$ leads to a
complicated behavior of the phase difference $\varphi(t)$ in time.

In several papers, \cite{Balatsky,Shumeiko,Fogel} dynamics of SmS
(superconductor-magnetic impurity-superconductor) JJs have been studied, where m
stands for a magnetic impurity. Interaction between tunneling Cooper pairs and the
magnetic moment of the impurity not only changes the current-phase relation
$I_{J}(\varphi)$ but also results in interesting dynamics of the magnetic moment.

The most interesting dynamic effects arise in tunnel JJs with a
ferromagnetic layer (or layers). In this case, the interaction between the
magnetization in F and the Josephson current is realized in the simplest
way. As is well known, even a weak in-plane magnetic field strongly affects
the Josephson current $I_{J}(\varphi )$. In case of JJs of the SIFS or SFIFS
type, such a magnetic field is produced by the F layer itself. Therefore any
perturbations of the magnetization vector $\mathbf{M}$ change the current $I_{J}(\varphi)$
and in addition the Meissner currents in the superconducting leads change
the orientation of the $\mathbf{M}$ vector.

In absence of the F layer, Josephson plasma waves can propagate in
SIS junctions and their spectrum is \cite{KulikJanson,Likharev,Barone}:
$\omega^{2}=\Omega _{J}^{2}+k^{2}v_{J}^{2}$, where $\Omega_{J}$ is the
Josephson ``plasma'' frequency and $v_{J}$ is the velocity of Swihart
waves. On the other hand, in the F film, spin waves can be excited with
the spectrum: $\omega ^{2}=\Omega _{M}^{2}(1+k^{2}l_{M}^{2})^{2}$, where $%
\Omega _{M}$ is the magnetic resonance frequency and $l_{M}$ is a
``magnetic'' length \cite{LL_ED}. If $\Omega_{J} < \Omega_{M}$, then these
dispersion curves cross (usually $l_{M}^{2}\ll l_{J}^{2}\equiv
v_{J}^{2}/\Omega _{J}^{2}$), and the interaction between magnetization
and Josephson currents leads to a coupling between Josephson ``plasma''
and spin waves and to a repulsion of the corresponding dispersion
``terms.'' The coupling between magnetic and superconducting oscillations
can be observed by studying the $I$-$V$ characteristics of the junction in the
presence of a weak external magnetic field. In this case, the so-called Fiske
steps arise on the $I$-$V$ curve, but their particular positions and form depend on
parameters characterizing the magnetic system. New peaks related to magnetic
resonances appear on the current-voltage characteristics (CVC). These results
have been obtained in a short paper by two of us \cite{VE}.

In the current paper, we study dynamic phenomena in the same systems (SIFS or
SFIFS JJs) as in Ref.~\onlinecite{VE}. However, we present in more detail the derivation of
equations describing the dynamics of the coupled magnetic and
superconducting systems (see Sec.~\ref{section-Model}). In particular, we
formulate conditions (frequency range) under which these equations are
valid. As in Ref.~\onlinecite{VE}, we analyze Fiske steps in SFIFS
junctions, but the CVC will be presented for a wider range of parameters of
these junctions. The CVC will be displayed not only for junctions with $%
L=l_{J}$ as it was done in Ref.~\onlinecite{VE}, but for junctions longer or
shorter than the Josephson length ($L < l_{J}$). The latter case corresponds
to the current experimental situation.

The coupled magneto-plasma modes will also be discussed in more detail
(see Sec.~\ref{section-CoupledModes}). Finally, in Sec.~\ref{section-FR}, we present a formula for the power
absorption $\mathcal{P}$ in SFIFS junctions when a weak ac in-plane
magnetic field is applied, that is, we study the ferromagnetic resonance in
the system. This formula drastically differs from the known formula for
ferromagnetic resonance in an isolated F film. In particular, it describes
plasma resonances in tunnel JJs, which also occur in absence of
the F film. The frequency dependence of $\mathcal{P}$ will be presented for
various system parameters. In Sec.~\ref{section-Discussion}, we discuss the obtained
results and analyze possibilities to observe the predicted effects in experiments.

\begin{figure}[t]
\includegraphics{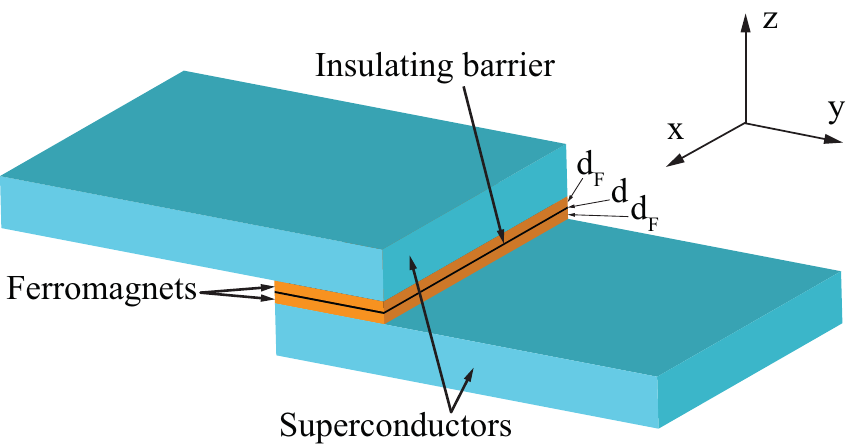}
\caption{(Color online) Schematic construction of a SFIFS junction of the
\textquotedblleft overlap\textquotedblright geometry.}
\label{fig1-setup}
\end{figure}

\section{Model and Basic Equations}
\label{section-Model}

We consider a planar SFIFS junction of the ``overlap'' geometry as shown schematically in
Fig.~\ref{fig1-setup} (the results obtained are also applicable to an SIFS junction).
Our aim is to generalize the equation for the phase difference $\varphi$
between the superconducting layers describing the static and dynamic
properties of an SIS JJ to the case of SFIFS JJs.

This equation reads \cite{Josephson1,Josephson2,KulikJanson,Likharev,Barone}

\begin{equation}
\Omega_J^{-2} \left( \frac{\partial^2\varphi}{\partial t^2} + \gamma_R \frac{%
\partial \varphi}{\partial t} \right) - l_J^2 \mathbf{\nabla}_{\perp}^2
\varphi + \sin(\varphi) = \eta,  \label{eq-varphiDynamics}
\end{equation}
where $\Omega_J = (2ej_c/C_{\Box}\hbar)^{1/2}$ is the Josephson
``plasma'' frequency, $\gamma_R = (R_{\Box}
C_{\Box})^{-1}$, $C_{\Box} = \epsilon/4\pi d$, and $R_{\Box}$ are the
capacitance and resistance of the junction per unit area, respectively, $d$
is the thickness of the insulating layer, $l_J^2 = v_J^2/\Omega_J^2$, $v_J =
c \sqrt{d/2\epsilon \lambda_L}$ is the plasma wave propagation velocity
(Swihart waves), $\lambda_L$ is the London penetration depth, and $\mathbf{%
\nabla}_{\perp}$ represents the tangential or in-plane gradient with respect
to the interfaces in the $x$-$y$ plane.

We single out the term on the right-hand side of Eq.~(\ref{eq-varphiDynamics}%
), $\eta = j/j_c$, which describes the normalized bias current through the
junction. Although it may depend on $y$, the normalized current $\eta$ will
be considered as constant along the $y$ direction. Strictly speaking, this
is only true for ``overlap'' junctions \cite{Likharev,Barone} considered here in which the system geometry is arranged
in such a way that the intersection region of superconducting layers is
approximately one-dimensional. However, the form of Eq.~(\ref{eq-varphiDynamics})
is most convenient for analysis of CVC for the system
under consideration and, moreover, neglecting the $y$ dependence of
normalized current $\eta$ does not change qualitatively the final results.
The critical current density $j_{c}$ is considered as a known quantity. It
was calculated in Refs.~\onlinecite{BVEcrcur,BB03,Vasenko11,Pugach11}.

The resistance $R_{\Box}$ depends on the voltage $V$ across the junction.
This dependence is especially strong in the case of tunnel SIS JJs if
the voltage $V$ is close to the energy gap $\Delta$. We assume that the
characteristic frequencies ($\Omega_{J}$ and $\Omega_{M}$) are smaller than $%
\Delta/\hbar$. In addition, we are interested in the form of the CVC at
voltages $V$ close to $\hbar\Omega_{J}/2e, \hbar\Omega_{M}/2e$, where $R_{\Box}
$ and, therefore, $\gamma_{R}$ can be regarded as constant. Of course, the
overall form of the CVC will be modified as a direct consequence of the
voltage-dependent damping coefficient $\gamma_{R}(V)$.

We consider planar JJs of the SFIFS, SFIS, or SFS type and assume that the
layer separating the two superconductors is characterized by the magnetic
susceptibility $\chi(\omega, k)$. In particular, this layer may be a
magnetic insulator or metallic ferromagnet. The derivation of an equation
for the phase difference $\varphi$ in SFIFS junctions is quite similar to
that in the case of tunnel SIS junctions \cite{Kulik,Scalapino,Likharev,Barone}. We
assume that there is no magnetic field normal to the interfaces in the
superconductors or, in other words, no Abrikosov vortices pierce the
superconducting films, and the lateral dimensions $L_{x,y}$ are much larger
than the thickness $d_F$ of the F layers and the Josephson penetration depth
$\lambda_L$. Since the normal component of the magnetic induction $B_z$ is
continuous at the superconductor-ferromagnet (SF) interfaces, it also
vanishes in the ferromagnetic layers and, hence, according to $B_z = H_z + 4\pi
M_z$ one has $H_z = -4\pi M_z$ in the F films. In order to find the relation
between the magnetic field $H$ in the superconductor (note that in the S
layers $H$ coincides with the magnetic induction $B$) and the phase
difference $\varphi$, we express the tangential component of the current
density in the S film $\mathbf{j}_{\perp} \equiv j_x \mathbf{n}_x + j_y
\mathbf{n}_y$ using the vector potential $A_{\perp}$ ($\mathbf{n}_z \times
\partial \mathbf{A}/\partial z = \mathbf{B}_{\perp}$) and the tangential
gradient of the phase in the superconductor $\nabla_{\perp}\chi$ as

\begin{equation}
\mathbf{j}_{\perp} = \frac{c}{4\pi \lambda_L^2} \left(1 + \gamma_{qp}\right)
\left(
-\mathbf{A}_{\perp} - \frac{\Phi_0}{2\pi}\mathbf{\nabla}_{\perp}\chi
\right),
\label{eq-TangentialCurrentDensity}
\end{equation}
where $\gamma_{qp}(\omega) = 4\pi i\omega \sigma(\omega) \lambda_L^2/c^2$ is
a damping parameter describing effects of quasiparticles on the supercurrent
and $\Phi_0 = hc/2e>0$ is the magnetic flux quantum. The parameter $%
\gamma_{qp}$ is very small for not very high frequencies because the
frequency $c/\lambda_L$ is very large. For example, taking $\lambda_L =
5\cdot 10^{-6}\mathrm{cm}$ we obtain $c/\lambda_L = 0.6 \cdot 10^{16}\mathrm{%
s}^{-1}$, which actually allows us to omit the parameter $\gamma_{qp}$.

Writing Eq.~(\ref{eq-TangentialCurrentDensity}) we imply a local relation
between the tangential current density $\mathbf{j}_{\perp}$ and the gauge
invariant quantity in brackets, which is legitimate in the limit $%
k\lambda_{L} \ll 1$, where $k$ is the modulus of the in-plane wave vector of
perturbations. Subtracting the expressions for the current density, Eq.~(\ref%
{eq-TangentialCurrentDensity}), written for the right and left
superconductors from each other we find the change of the tangential current
density $[\mathbf{j}_{\perp}]=\mathbf{j}_{\perp}(\widetilde{d}_F/2) - \mathbf{j}_{\perp}(-\widetilde{d}_F/2)$
across the junction

\begin{equation}
[\mathbf{j}_{\perp}]
=
\frac{c}{4\pi \lambda_L^2} \left(1 +
\gamma_{qp}\right)
\left. \left( \widetilde{d}_F \left\{ \mathbf{n}_z \times \mathbf{B%
}_{\perp} \right\} - \frac{\Phi_0}{2\pi} \nabla_{\perp} \varphi \right)
\right|_{\frac{\widetilde{d}_F}{2}},
\label{eq-CurrentDensityJump}
\end{equation}
where $\widetilde{d}_F = d_F$ in the case of an SIFS or SFS junction and $%
\widetilde{d}_F = 2d_F$ in the case of an SFIFS junction. The parameter $d_F$ is the
thickness of the F film, which is assumed to be smaller than the London penetration
length $\lambda_L$, and for any quantity $Q$, we denote the difference $Q\big|%
_{S(R)} - Q\big|_{S(L)}$ by $[Q]$, where $S(R)$ and $S(L)$ are the right and left
superconductors, respectively.

The assumption $d_{F} \ll \lambda_L$ allows one to neglect the change of $%
\mathbf{A}_{\perp}$ along the $z$ direction caused by Meissner currents in
the F layer and to write the change of the vector potential $\mathbf{A}%
_{\perp}$ in the form $[\mathbf{A}_{\perp}] = \widetilde{d}_F(\mathbf{n}_z
\times \mathbf{B}_{\perp})$ with $\mathbf{B}_{\perp} = 4\pi \mathbf{M}%
_{\perp} + \mathbf{H}_{\perp}$. The field $\mathbf{H}_{\perp}$ is
approximately the same to the right and to the left from the SF interfaces
and does not contribute to the jump of the tangential current density $[\mathbf{j}_{\perp}]$.
The Meissner currents in the F layers and, therefore, the
variation of $\mathbf{H}_{\perp}$ there are much smaller than in the
superconductors for the following reason. The total screening Meissner
current $I_{\mathrm{Scr}}$ in the F layer is proportional to $%
\lambda_{LF}^{-2} \widetilde{d}_F \mathbf{A}$, where the inverse London
penetration depth $\lambda_{LF}^{-1}$ is proportional to the density of
Cooper pairs, $\lambda_{LF}^{-2} \sim n_{SF}$, and, thus, is much smaller
than $\lambda_L^{-2}$. The phase difference $\varphi$ between the two S
layers has the (gauge-invariant) definition:

\begin{equation}
\varphi = [\chi] + \frac{2e}{\hbar c} \int_{S(L)}^{S(R)} dz\, A_z,
\label{eq-VarphiGaugeInvariant}
\end{equation}
and completely describes the JJ because we choose a gauge with $A_z = 0$ and
$[\chi]=\chi(\widetilde{d}_F/2) - \chi(-\widetilde{d}_F/2)$.

Equation~(\ref{eq-CurrentDensityJump}) determines the boundary conditions of
the London equation in the superconductors. Indeed, considering the Maxwell
equation at the points $z = \pm z_{\mathrm{SF}} \approx \pm \widetilde{d}_F/2$,

\begin{equation}
\mathbf{\nabla} \times \mathbf{B}_{\perp} \Big|_{\pm \widetilde{d}_F/2} =
\frac{4\pi }{c} \mathbf{j}_{\perp} \Big|_{\pm \widetilde{d}_F/2},
\label{eq-MaxwellRotB}
\end{equation}
where $z_{\mathrm{SF}}$ denotes the coordinate of the right SF interface, we obtain
by successively taking the cross product with $\mathbf{n}_z$ in both sides
and subtracting the two equations from each other

\begin{equation}
- \frac{\partial \mathbf{B}_{\perp}}{\partial z} \Big|_{\widetilde{d}_F/2} =
\frac{2\pi}{c} \mathbf{n}_z \times [\mathbf{j}_{\perp}].
\label{eq-MaxwellRotBjump}
\end{equation}
Here, we used the relation

\begin{equation}
\left. \frac{\partial \mathbf{B}_{\perp}}{\partial z} \right|_{\widetilde{d}%
_F/2} = -\left. \frac{\partial \mathbf{B}_{\perp}}{\partial z} \right|_{-%
\widetilde{d}_F/2}  \label{eq-SymmetryRelationB}
\end{equation}
taking into account the symmetry of the SFIFS system. Recalling that the
magnetic field component $B_{z}$ normal to the interfaces is assumed to be
zero in the S layers and considering only the $z$ dependence of $\mathbf{B}%
_{\perp}$, we have to solve in the superconductors the equation

\begin{equation}
\frac{\partial^2 \mathbf{B}_{\perp}}{\partial z^2} - \kappa^2 \mathbf{B}%
_{\perp} = 0  \label{eq-SolveBinSlayer}
\end{equation}
with $\kappa^2 = \lambda_L^{-2} \left( 1+\gamma_{qp} \right) \approx
\lambda_L^{-2}$. The solution reads for $|z|~>~\widetilde{d}_F/2$,

\begin{equation}
\mathbf{B}_{\perp}(z) = \mathbf{B}_{\perp}\left( \widetilde{d}_F/2 \right)
\exp\left\{-\frac{|z|-\widetilde{d}_F/2}{\lambda_L} \right\}.
\label{eq-BinSlayer}
\end{equation}
Inserting this expression for $\mathbf{B}_{\perp}$ into Eq.~(\ref%
{eq-MaxwellRotBjump}) we obtain by use of Eq.~(\ref{eq-CurrentDensityJump})

\begin{equation}
\mathbf{B}_{\perp}\left( \tfrac{\widetilde{d}_F}{2} \right) = -\frac{\Phi_0}{%
4\pi \widetilde{\lambda}_L} \left( \mathbf{n}_z \times \nabla_{\perp}
\varphi \right) - \frac{2\pi \widetilde{d}_F}{\widetilde{\lambda}_L} \mathbf{%
M}_{\perp}\Big|_{\widetilde{d}_F/2},  \label{eq-BinSlayerPrefactor}
\end{equation}
where we have set $\widetilde{\lambda}_L = \lambda_L + \widetilde{d}_F/2$.
The magnetic field $B$ decays exponentially with increasing $z$ provided the
thickness of the S layers exceeds the London penetration length $\lambda_L$.

In order to obtain an equation for the phase difference $\varphi$ of the
superconductors we use the Maxwell equation
$(\nabla~\times~\mathbf{H})_z~-~c^{-1}~\partial~D_z/\partial~t~=~(4\pi/c)~j_z$ and the
standard expression for the Josephson current according to the
Stewart-McCumber model \cite{Stewart,McCumber}. This simple model [also known
as the resistively and capacitively shunted junction (RCSJ) model] provides
a good description of the CVC of a real JJ, although effects due to finite
dimensions of the contacts and nonlinearities of the quasiparticle current
are neglected. Using the Josephson relation

\begin{equation}
\frac{\partial \varphi}{\partial t} = -\frac{2eV}{\hbar}
\label{eq-JosephsonRelation}
\end{equation}
and the standard expression for the Josephson current we obtain within this
model

\begin{equation}
\frac{c}{4\pi} \left( \nabla \times \mathbf{H} \right)_z = -\frac{\hbar
C_{\Box}}{2e} \frac{\partial^2 \varphi}{\partial t^2} - \frac{\hbar}{%
2eR_{\Box}} \frac{\partial \varphi}{\partial t} - j_c \sin(\varphi) + j.
\label{eq-MaxwellAndStandardJJcurrent}
\end{equation}
Finally, with the help of Eq.~(\ref{eq-BinSlayerPrefactor}) and taking
into account that in the S layers $B=H$,

\begin{eqnarray}
&& \Omega_J^{-2} \left( \frac{\partial^2 \varphi}{\partial t^2} + \gamma_R
\frac{\partial \varphi}{\partial t} \right) - l_J^2 \nabla_{\perp}^2 \varphi
+ \sin(\varphi) =  \notag \\
&& \qquad\qquad\qquad\qquad = \eta + \frac{c\widetilde{d}_F}{2\widetilde{%
\lambda}_L j_c} \left( \nabla \times \mathbf{M}_{\perp} \right)_z
\label{eq-VarphiDynamicsAndM}
\end{eqnarray}
where here, too, $\Omega_J = (2ej_{c}/C_{\Box} \hbar)^{1/2}$ is the
Josephson ``plasma'' frequency, $\gamma_R =
(R_{\Box}C_{\Box})^{-1}$, $C_{\Box}(\omega) = \epsilon(\omega)/4\pi d$ and $%
R_{\Box}(\omega)$ are the capacitance and resistance of the junction per
unit area, respectively, $d$ is the thickness of the insulating layer, $%
l_J^2 = v_J^2/\Omega_J^2$, $v_J = c\sqrt{d/2\epsilon \widetilde{\lambda}_L}$
is the plasma wave propagation velocity (Swihart waves), and $\eta = j/j_c
$ is the normalized bias current through the junction. The capacitance $C_{\Box}$
and the resistance $R_{\Box}$ of the junction may depend on frequency $\omega$ (in
the Fourier representation). A simpler equation for the phase difference $%
\varphi$ in the stationary case has been reported previously in Ref. %
\onlinecite{AV}. In a general, non-stationary case, this equation was derived
in Ref. \onlinecite{VE}. Note that a slightly different approach for the
study of dynamic processes in SFS junctions was used in a recent paper \cite%
{Maekawa}. In particular, Eq.~(\ref{eq-VarphiDynamicsAndM}) can be easily
derived from Eqs.~(A3)--(A6) of this work.

In order to obtain a closed set
of equations for the phase difference $\varphi$ of the superconductors and
the magnetization $\mathbf{M}_{\perp}$ of the ferromagnetic layer, we need
to use a dynamic equation for $\mathbf{M}_{\perp}$ as well.

The dynamics of the magnetization $\mathbf{M}$ in the F layer is described
by the well-known Landau-Lifshitz-Gilbert (LLG) equation (see, e.g., Refs.~%
\onlinecite{LL_ED,Aharoni}), which allows one to describe the temporal
development of $\mathbf{M}$ in an effective magnetic field $\mathbf{H}_{%
\mathrm{eff}}$ including all internal and external contributions.

We decompose the magnetization vector $\mathbf{M}$ according to $\mathbf{M}
= M_0 \mathbf{n}_e + \mathbf{m}$, where the unit vector $\mathbf{n}_e$
denotes the easy axis direction and $\mathbf{m} \perp \mathbf{n}_e$ is the
dynamic part which evolves in time as described by the LLG equation.
Assuming that in equilibrium the magnetization coincides with the static
part along the easy axis, i.e. $M_0 \approx |\mathbf{M}| \gg |\mathbf{m}|$,
and using $B_z = 0$, we obtain

\begin{eqnarray}
\frac{\partial \mathbf{m}}{\partial t} &=& -4\pi \alpha M_{\mathrm{eff}}
\left( 1-\widetilde{l}_M^{\,2} \nabla_{\perp}^2\right)
\left( \mathbf{M}\times \mathbf{m}\right) +
\notag
\\
&&
+ M_{\mathrm{eff}} \mathbf{M} \times \mathbf{B}_{\perp}
+ \frac{\gamma_M}{|\mathbf{M}|}\mathbf{M} \times \frac{\partial \mathbf{m}}{\partial t},
\label{eq-LLG}
\end{eqnarray}
where $M_{\mathrm{eff}} = g|e|/2mc$, $g<0$ is the gyromagnetic factor, $\alpha$ is a
parameter related to the anisotropy constant \cite{LL_ED}, $\widetilde{l}_M$
is a characteristic length related to spin waves, and $\gamma_M$ is the
dimensionless Gilbert damping constant.

We further neglect the Gilbert damping term ($\gamma_M = 0$), align the easy
axis along the $z$ direction ($e \equiv z$), and substitute $\mathbf{B}%
_{\perp F} = 4\pi \mathbf{M}_{\perp} + \mathbf{H}_{\perp}$ $(\mathbf{M}%
_{\perp} \equiv \mathbf{m})$ into Eq.~(\ref{eq-LLG}), where $\mathbf{B}%
_{\perp F}$ is the magnetic induction in the F layer and $\mathbf{H}_{\perp}$
is the magnetic field, which is assumed to be independent of the $z$
coordinate (screening effects in the F layer are negligible). The field $%
\mathbf{H}_{\perp}$ is continuous across the SF interface, i.e., $\mathbf{H}%
_{\perp} = \mathbf{B}_{\perp}\left(z \rightarrow \widetilde{d}_F/2\right)$,
and is given by Eq.~(\ref{eq-BinSlayerPrefactor}).

Finally, we obtain

\begin{eqnarray}
\frac{\partial \mathbf{m}}{\partial t}
&=&
\Omega_M \left[ \left( 1+s-l_M^2 \nabla_{\perp}^2 \right) \frac{\mathbf{M} \times \mathbf{m}}{M_0}
-
\right.
\nonumber
\\
&&
\phantom{\Omega_M \Big[}
\left.
\qquad
-\frac{\Phi_0}{(4\pi)^2 (\alpha - 1) \widetilde{\lambda}_L} \nabla_{\perp} \varphi \right],
\label{eq-LLGvarhiCoupled}
\end{eqnarray}
where $\Omega_M = 4\pi(\alpha-1) |M_{\mathrm{eff}}| M_0$ is the
resonance frequency of magnetic moment precession ($\alpha > 1$), $s =
\widetilde{d}_F/[2 (\alpha-1) \widetilde{\lambda}_L]$, $l_M^{\,2} = [\alpha/(\alpha-1)]
\widetilde{l}_M^2$.

Equations~(\ref{eq-VarphiDynamicsAndM}) and (\ref{eq-LLGvarhiCoupled}) fully
describe different dynamical processes in the junctions under consideration.
Note that the Josephson current is coupled to the magnetization through the
spatial derivative of the phase difference $\mathbf{\nabla}_{\perp}\varphi$
[the last term on the right-hand side of Eq.~(\ref{eq-LLGvarhiCoupled})].
Therefore, in a spatially homogeneous case there is no coupling between the
Josephson effect and dynamics of the magnetization.

\section{Fiske steps}

\label{section-LongFiske}

In this section, we consider a SFIFS Josephson junction in a weak external magnetic field $H_{%
\mathrm{ext}}$ assuming that it is constant in space and time and is
directed parallel to the interfaces along the $y$ direction. As is well
known, in this case so-called Fiske steps arise on the CVC due to excitation
of eigenmodes in the junction. The phase difference $\varphi(x,t)$ depends
on the $x$ coordinate and, therefore, dynamics of the magnetic and superfluid
systems are coupled together. We consider the case when the magnetization
vector in the stationary state is directed perpendicular to the SF
interfaces, i.e. $\mathbf{M}_0 = M_0 \mathbf{n}_z$ and $\mathbf{H}_0 = -4\pi
\mathbf{M}_0$. As the typical values for the magnitude of the stationary
magnetization $M_0$ are hundreds of Gau\ss\, and the small external magnetic
field $H_{\mathrm{ext}}$ is of the order of a few Gau\ss, one can neglect
the in-plane magnetization $M_y=-H_{\mathrm{ext}}/(4\pi)$ compared to $M_0$.
The resulting precessional motion of the magnetization $\mathbf{M}$ in
presence of a current through the JJ implies that the in-plane components $%
\mathbf{m} \perp \mathbf{n}_z$ of $\mathbf{M}$ are excited. Therefore, we
represent $\mathbf{M}$ as $\mathbf{M}(x,t) = \mathbf{M}_0 + \mathbf{m}(x,t)$.
Components $m_{x,y}$ are easily found from Eq.~(\ref{eq-LLGvarhiCoupled}):

\begin{eqnarray}
m_y
&=&
\frac{\Omega_M(1+s)}{i\omega} m_x
\nonumber
\\
&=&
\frac{1}{(1+s)\mathcal{L}_{\omega F}} \frac{\Phi_0}{(4\pi)^2 (\alpha-1) \widetilde{\lambda}_L}
\frac{\partial \varphi}{\partial x},
\quad
\label{eq-InplaneComponentsM}
\\
\mathcal{L}_{\omega F}
&=&
\frac{\omega(\omega-i\gamma_M)}{\Omega_M^2(1+s)^2}-1.
\label{eq-DefLomegaF}
\end{eqnarray}
Equations~(\ref{eq-InplaneComponentsM}), (\ref{eq-DefLomegaF}) are written
under the assumption that all relevant quantities depend on time as $%
\exp(i\omega t)$ and, what is more important, spatial derivatives in the
equation for $\mathbf{m}(x,t)$ are neglected. The latter assumption is
justified provided the magnetic length $l_M$ is much shorter than the
Josephson length $l_J$: $l_M \ll l_J$. It is not difficult to analyze a more general case of
arbitrary relation between $l_{M}$ and $l_{J}$, but the corresponding
formulas become too cumbersome. Substituting Eq.~(\ref{eq-InplaneComponentsM})
into Eq.~(\ref{eq-VarphiDynamicsAndM}) we obtain

\begin{eqnarray}
- \left[ \frac{\omega (\omega - i\gamma_R)}{\Omega_J^2} +
\widetilde{l}_J^{\,2}(\omega) \frac{\partial^2}{\partial x^2} \right] \varphi(x,\omega)
+
\qquad\qquad
&&
\label{eq-FinalODEvarphi}
\\
+ \mathcal{F}\{ \sin(\varphi) \}(x,\omega)
&=&
\eta,
\nonumber
\end{eqnarray}
where $\mathcal{F}\{ \sin(\varphi) \}(x,\omega)$ is the Fourier transform of
$\sin[\varphi(x,t)]$ with respect to time $t$ and

\begin{equation}
\widetilde{l}_J(\omega) = l_J \left[ 1 + \frac{s}{(1+s)\mathcal{L}_{\omega F}} \right]^{1/2}
\label{eq-renormalizedLength}
\end{equation}
is a renormalized Josephson length containing $\mathcal{L}_{\omega F}$ and,
therefore, depending on frequency $\omega$. Equation~(\ref{eq-FinalODEvarphi}) is the favored
generalization of Eq.~(\ref{eq-varphiDynamics}) for SFIFS junctions.

In order to find the CVC, we represent the phase difference $\varphi$ of the
superconducting layers in the form $\varphi = \varphi_0(x,t) +
\psi (x,t)$ (see Ref.~\onlinecite{Scalapino}). The first term is given by $%
\varphi_0(x,t) = \kappa_H x + \Omega_V t$ with $\kappa_H = 4\pi \widetilde{%
\lambda}_L H_{\mathrm{ext}}/\Phi_0$ [see Eq.~(\ref{eq-BinSlayerPrefactor})]
and $\Omega_V = 2eV/\hbar$. The function $\psi(x,t)$ is assumed to be small
allowing us to linearize Eq.~(\ref{eq-VarphiDynamicsAndM}) with respect to
$\psi$:

\begin{eqnarray}
-\mathit{\widehat{P}} \{ \psi \}(x,t) &=& \sin \left[ \varphi_0(x,t) \right]
\label{eq-ODEpsi} \\
&=& \sin(\Omega_V t) \cos(\kappa_H x) + \cos(\Omega_V t) \sin(\kappa_H x)
\notag
\end{eqnarray}
where the operator $\mathit{\widehat{P}}$ is defined as

\begin{equation}
\mathit{\widehat{P}} = \Omega_J^{-2} \left( \frac{\partial^2}{\partial t^2}
+ \gamma_R \frac{\partial}{\partial t} \right) - \widetilde{l}%
_J^{\,2}(\Omega_V) \frac{\partial^2}{\partial x^2}.  \label{eq-defOperatorP}
\end{equation}
The current correction $\delta\eta$ to the dc current $\eta_0 =
(2eV/\hbar)/\Omega_J = \Omega_V/\Omega_J$ is given by

\begin{equation}
\delta\eta = \big\langle \psi(x,t) \cos \left[\varphi_0(x,t)\right] %
\big\rangle,  \label{eq-CurrentEtaPsiAvg}
\end{equation}
where the angular brackets denote the average with respect to space and time.

Equation~(\ref{eq-CurrentEtaPsiAvg}) determines the constant normalized
current through the junction as a function of voltage $V$, which gives a current-voltage
(I-V) curve. Equation~(\ref{eq-ODEpsi}) contains parts oscillating in space
and time. It should be solved taking into account the boundary conditions %
\cite{Scalapino,Kulik,Likharev,Barone}

\begin{equation}
\left. \frac{\partial\psi}{\partial x} \right|_{x=\pm L}=0,
\label{eq-BoundaryConditionPsi}
\end{equation}
where $L$ denotes the length of the junction along the $x$ direction.
The right-hand side of Eq.~(\ref{eq-ODEpsi}) can be written in the form $%
\mathrm{Im}\{\exp(i\Omega_V t)[\cos(\kappa_Hx) + i\sin(\kappa_H x)]\}$ and,
therefore, the solution of Eq.~(\ref{eq-ODEpsi}) can be written as $%
\psi(x,t) = \mathrm{Im}\{\exp(i\Omega_V t)\psi_1(x)\}$, where the function $%
\psi_1(x)$ obeys the equation

\begin{equation}
-\mathit{\widehat{P}}_{\Omega} \{ \psi_1 \}(x) = \cos(\kappa_H x) +
i\sin(\kappa_H x)  \label{21}
\end{equation}
with the boundary condition Eq.~(\ref{eq-BoundaryConditionPsi}).

\begin{figure*}[t!]
\includegraphics[scale=0.25]{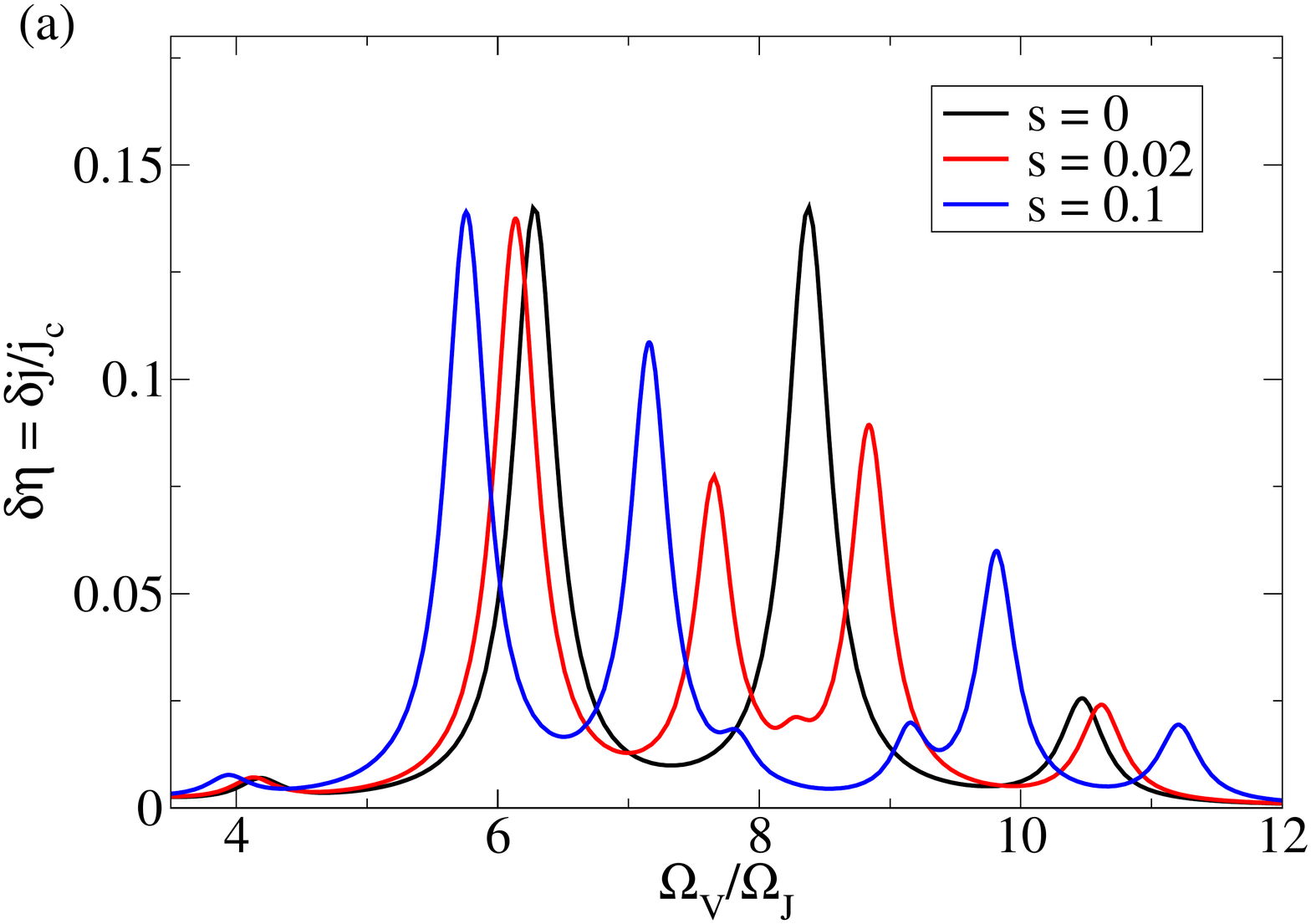}\qquad %
\includegraphics[scale=0.25]{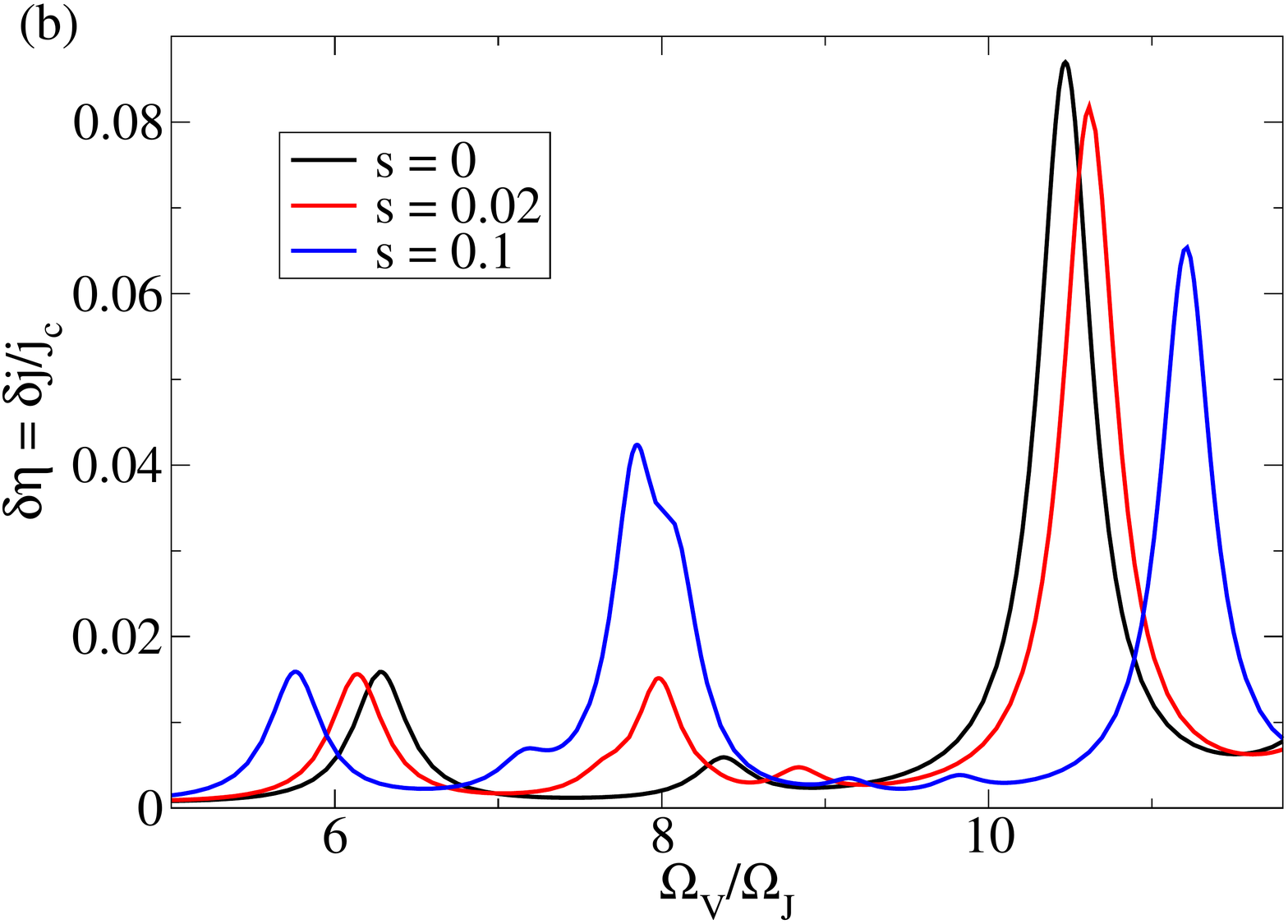} %
\includegraphics[scale=0.25]{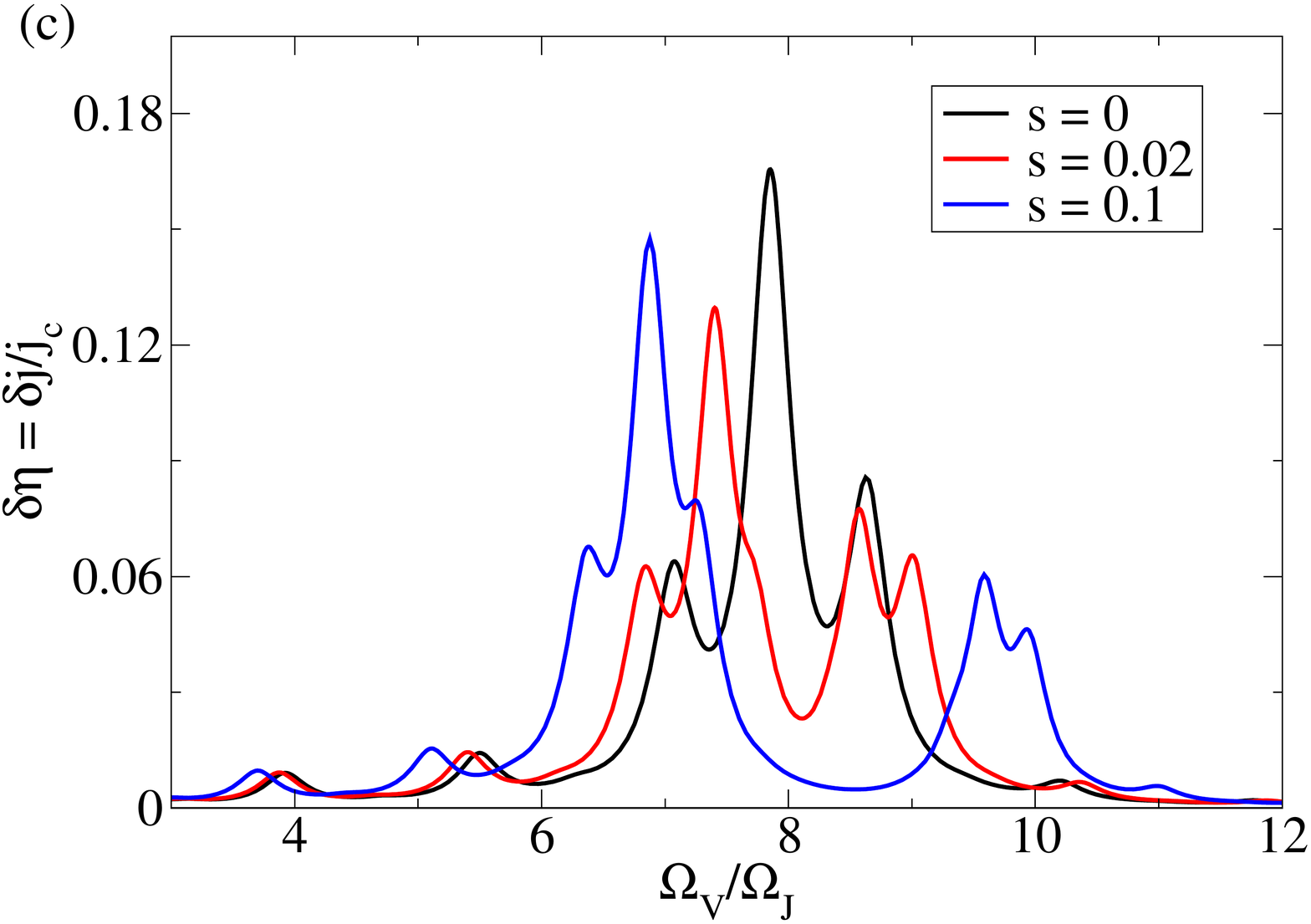}\qquad %
\includegraphics[scale=0.25]{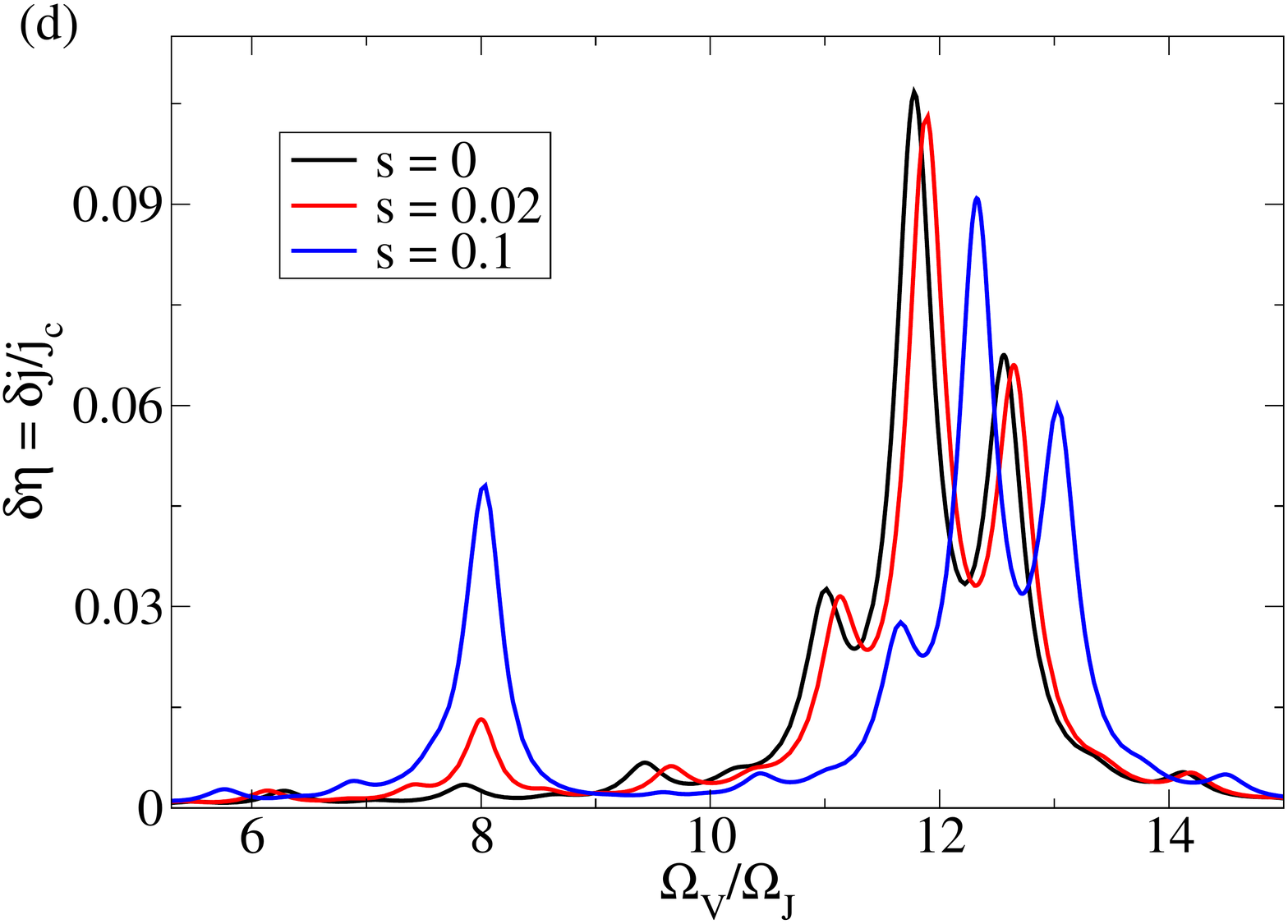}
\caption{(Color online) Correction to the $I$-$V$ characteristics of an SFIFS JJ
in a weak external magnetic field due to interaction of Josephson
oscillations and spin-wave modes. The correction is plotted as a function of
normalized voltage $V_{\mathrm{norm.}} = \Omega_V/\Omega_J$ for different
values of the parameter $s$. The figures are presented for the following
parameters: (a) $\Omega_M/\Omega_J = \kappa_H l_J = 8$ and $L/l_J = 0.75$,
(b) $\Omega_M/\Omega_J = 8, \protect\kappa_H l_J = 12, L/l_J = 0.75$; (c) $%
\Omega_M/\Omega_J = \protect\kappa_H l_J = 8, L/l_J = 2$; (d) $%
\Omega_M/\Omega_J = 8, \protect\kappa_H l_J = 12, L/l_J = 2$. The damping
coefficients are $\protect\gamma_R/\Omega_J = 0.4, \protect\gamma_M/\Omega_J
= 0.3$.}
\label{fig-LongJJcurrentCorrection1}
\end{figure*}

The operator $\mathit{\widehat{P}}_{\Omega}$ coincides with $\mathit{%
\widehat{P}}$ after replacing $\partial/\partial t$ by $i\Omega_V$. The
solution can be easily found and equals

\begin{eqnarray}
\psi_1 &=& \frac{1}{\mathit{P}_{\Omega}(V,H)}\Big\{ \cos(\kappa_H x) +
C\cos(\kappa_V x) + i \sin(\kappa_H x) +  \notag \\
&& \phantom{\frac{1}{\mathit{\widehat{P}}_{\Omega}(V,H)}\{} +
iS\sin(\kappa_V x)\Big\}  \label{eq-ResultForPsi}
\end{eqnarray}
where $\mathit{P}_{\Omega}(V,H) = a^2 - \widetilde{l}_J^{\,2}(\Omega_V)%
\kappa_H^2$, $a^2 = \Omega_J^{-2}(\Omega_V^2 - i\gamma_R \Omega_V)$ and

\begin{equation}
C = -\frac{\theta_H}{\theta_V} \frac{\sin \theta_H}{\sin \theta_V}, \quad S
= -\frac{\theta_H}{\theta_V} \frac{\cos \theta_H}{\cos \theta_V}
\label{eq-DefParametersOfPsi}
\end{equation}
with $\theta_H = \kappa_H L,\, \theta_V = \kappa_V L,\, \kappa_V^2 = a^2%
\widetilde{l}_J^{\,-2}(\Omega_V)$. Substituting the function $\psi(x,t)$ expressed
through $\psi_1(x)$ into Eq.~(\ref{eq-CurrentEtaPsiAvg}), we find the dependence,
$\delta\eta(V) \equiv \delta j(V)/j_c$,

\begin{eqnarray}
\delta\eta &=& \mathrm{Im} \left\{ \frac{1}{\mathit{P}_{\Omega}(V,H)} \left[
1 - \frac{\theta_H^2}{\theta_V \left(\theta_H^2 - \theta_V^2 \right)} \times
\right.\right.  \notag \\
&& \phantom{\mathrm{Im} \Big\{\frac{1}{\mathit{\widehat{P}}_{\Omega}(V,H)}}
\times \left.\left. \frac{\cos(2\theta_V) - \cos(2\theta_H)}{\sin(2\theta_V)}
\right] \right\}.  \label{eq-ResultForEta}
\end{eqnarray}
Since we assumed that the correction $\psi = \mathrm{Im} \left\{ \exp \left( i\Omega
_{V}t\right) \psi _{1}(x)\right\} $ to the phase difference $\varphi$ in the
superconducting layers is small, Eq.~(\ref{eq-ResultForEta}) is only valid
for normalized voltages $\Omega _{V}/\Omega _{J}>(\gamma _{R}/\Omega
_{J})^{-1}$. This can be seen from Eq.~(\ref{eq-ResultForPsi}) where one
should verify that the prefactor $P_{\Omega }^{-1}(V,H)$ is small.

Let us discuss the current results and compare them with those
obtained in Ref.~\onlinecite{VE}. The prefactor $P_{\Omega}^{-1}(V,H)$
in Eq.~(\ref{eq-ResultForEta}) contains the renormalized Josephson
length $\widetilde{l}_{J}$ defined in Eq.~(\ref{eq-renormalizedLength}),
which corresponds to the quantity $l_{V}$ of Ref.~\onlinecite{VE}.
The formulas for Fiske steps in Ref.~\onlinecite{VE} were given for small values
of the parameter $s$. If the parameter $s$ is not very small, one can reproduce the correct result by replacing there
$\Omega_{M}^{2} \rightarrow (1+s)\Omega_{M}^{2}$, i.e., Eq.~(\ref{eq-renormalizedLength}).
[Note that in the definition of $\Omega_{Ms}$, Eq.~(10) of Ref.~\onlinecite{VE}, there is a misprint.
The factor of two in the exponent at the right-hand side is missing so that the
correct formula reads $\Omega_{Ms}^{2} = \Omega_{M}^{2}(1+s)^{2}$.]
The modified dependence of the normalized Josephson length $\widetilde{l}_{J}$
on the parameter $s$ changes the form of the $I$-$V$ characteristics and reveals that the
effect of the ferromagnetic layer is much more pronounced compared to the results of Ref.~\onlinecite{VE}
even for small $s$ because the denominator in Eq.~(10) of Ref.~\onlinecite{VE} is very small at voltages
corresponding to peaks in the CVC and, therefore, is very sensitive to the parameter $s$.
Thus, we update the figures showing the dependence $\delta \eta(V_{\mathrm{norm.}})$
as a function of normalized voltage $V_{\mathrm{norm.}} = \Omega_V/\Omega_J$. Finally, we also present
$I$-$V$ characteristics for different values of normalized junction lengths
$L/l_{J}$ including those which correspond to the experimental
values of Ref.~\onlinecite{Wild} ($L/l_{J}<1$). As in Ref.~\onlinecite{VE},
for simplicity, we assume that the damping coefficient $\gamma _{R}$
is constant, i.e., it does not depend on voltage $V$.

\begin{figure}[t!]
\includegraphics[scale=0.25]{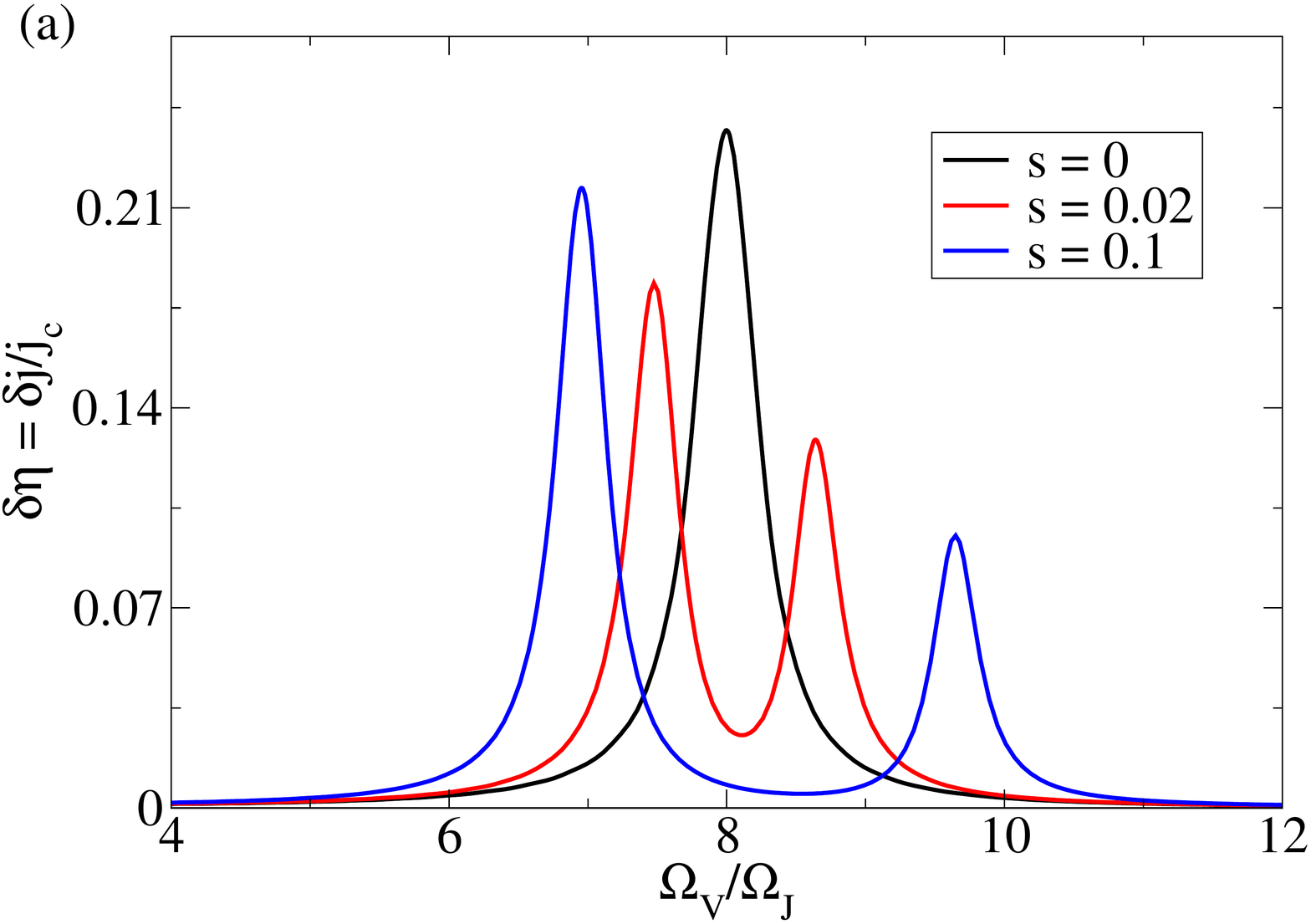} %
\includegraphics[scale=0.25]{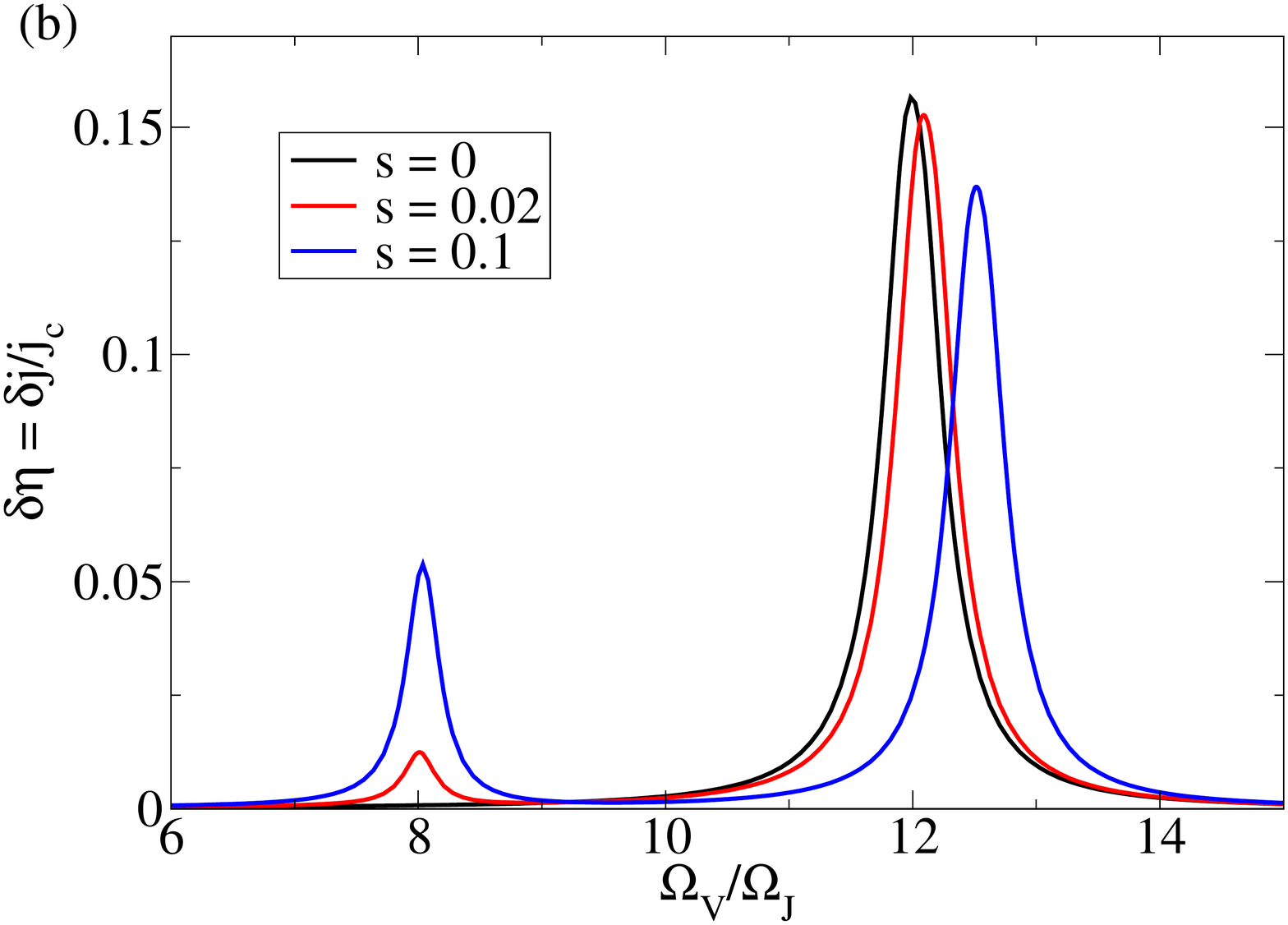}
\caption{(Color online) Current correction $\protect\delta\protect\eta$ as a
function of normalized voltage $V_{\mathrm{norm.}} = \Omega_V/\Omega_J$ for
long JJ with normalized junction length $L/l_J = 10$. The function $\protect%
\delta\protect\eta$ is displayed for (a) $\Omega_M/\Omega_J = \protect\kappa%
_H l_J = 8$; (b) $\Omega_M/\Omega_J = 8, \protect\kappa_H l_J = 12$ and for
different values of the parameter $s$. The damping coefficients are $\protect%
\gamma_R/\Omega_J = 0.4, \protect\gamma_M/\Omega_J = 0.3$.}
\label{fig-LongJJcurrentCorrection2}
\end{figure}

In Figs.~\ref{fig-LongJJcurrentCorrection1} and \ref{fig-LongJJcurrentCorrection2},
we plot the current correction $\delta \eta$ as a function of normalized voltage $V_{\mathrm{norm.}}$
for different values of the parameter $s=\widetilde{d}_{F}/(2 (\alpha-1) \widetilde{\lambda}_{L})$
and normalized junction length $L/l_{J}$. Taking into account the experimental values of $L$ and $l_{J}$ (see
Ref.~\onlinecite{Wild}), we display the current correction $\delta \eta $ for
short junctions with $L/l_{J}=0.75$ [see Figs.~\ref{fig-LongJJcurrentCorrection1}(a) and \ref{fig-LongJJcurrentCorrection1}(b)]
and, in addition, for longer junctions with $L/l_{J}=2$ [see Figs.~\ref{fig-LongJJcurrentCorrection1}(c) and \ref{fig-LongJJcurrentCorrection1}(d)] and $L/l_{J}=10$ [see Figs.~\ref{fig-LongJJcurrentCorrection2}(a) and
\ref{fig-LongJJcurrentCorrection2}(b)]. Black curves represent the limit $s\rightarrow 0$ where we have no
F layers in the system and the CVC correspond to ordinary Fiske steps. Due to the fact that in experiments,
only the strength of the external magnetic field can be varied, we display our result for different
values of the parameter $\kappa _{H}l_{J}\propto H_{\mathrm{ext}}$ keeping
all other system parameters such as $\Omega _{M}/\Omega _{J},L/l_{J}$, and $s$
constant.

The strongest influence of the ferromagnetic layers on the current-voltage
characteristics develops for external magnetic fields such that the
parameters $\kappa _{H}l_{J}$ and $\Omega _{M}/\Omega _{J}$ coincide. By
comparing Figs.~\ref{fig-LongJJcurrentCorrection1}(a) and \ref{fig-LongJJcurrentCorrection1}(b)
[or Figs.~\ref{fig-LongJJcurrentCorrection1}(c) and \ref{fig-LongJJcurrentCorrection1}(d), respectively] one can observe
that the change of the current correction is clearly recognizable for $%
\kappa _{H}l_{J}=\Omega _{M}/\Omega _{J}$ and nonzero $s$, while for $\kappa
_{H}l_{J}\neq \Omega _{M}/\Omega _{J}$, it only becomes pronounced for larger
values of $s$.

As can be seen from Figs.~\ref{fig-LongJJcurrentCorrection1}%
(a) and \ref{fig-LongJJcurrentCorrection1}(c) that the normalized junction length $L/l_{J}$ determines the form of
the CVC even in the case $s=0$, i.e., the number of Fiske steps close to the
normalized magnetic resonance frequency $\Omega _{M}/\Omega _{J}$ may vary
for different values of $L/l_{J}$. Provided for $s=0$ there appears a single
peak close to $\Omega _{M}/\Omega _{J}$, increasing the parameter $s$ leads
to a double splitting of the dominant peak. For even larger values of $s$,
the pair of peaks moves more and more apart from each other [see Fig.~\ref%
{fig-LongJJcurrentCorrection1}(a)]. A similar effect can be seen for a
larger number of Fiske steps close to $\Omega _{M}/\Omega _{J}$, e.g., Fig.~%
\ref{fig-LongJJcurrentCorrection1}(c) displays essentially two Fiske steps
in the vicinity of $\Omega _{M}/\Omega _{J}=8$ that both split up into two
peaks moving apart from each other with increasing $s$.

For distinct values of the parameters $\kappa _{H}l_{J}$ and $\Omega
_{M}/\Omega _{J}$ [see Figs.~\ref{fig-LongJJcurrentCorrection1}(b) and \ref{fig-LongJJcurrentCorrection1}(d)], there
also emerge additional peaks in the $I$-$V$ characteristics close to the
normalized magnetic resonance frequency, but the detailed impact of the F
layers on the CVC is not as obvious as is the case for $\kappa
_{H}l_{J}=\Omega _{M}/\Omega _{J}$. From Fig.~\ref%
{fig-LongJJcurrentCorrection1}(d), one can already conjecture that for long
junctions, the ferromagnetic layers simply induce a single additional peak
close to $\Omega _{M}/\Omega _{J}$. In Fig.~\ref{fig-LongJJcurrentCorrection2}, where the current correction $%
\delta \eta $ is shown for the limit of large values of $L/l_{J}$ ($%
L/l_{J}=10$), this feature becomes more apparent. For coinciding values of
the magnetic resonance frequency $\Omega _{M}/\Omega _{J}$ and the parameter
$\kappa _{H}l_{J}$ [see Fig.~\ref{fig-LongJJcurrentCorrection2}(a)], we find a
single peak for $s=0$ and a double peak for $s\neq 0$ in the vicinity of $%
\Omega _{M}/\Omega _{J}$. For $\Omega _{M}/\Omega _{J}\neq \kappa _{H}l_{J}$,
there emerges a single peak close to $\Omega _{M}/\Omega _{J}$ and $\kappa
_{H}l_{J}$, respectively, where the former is notably smaller in magnitude
[see Fig.~\ref{fig-LongJJcurrentCorrection2}(b)]. Below we also derive analytical
expressions for these peak positions.

Thus the presence of the F layers leads not only to a shift of the peaks in
the dependence $\delta\eta(V_{\mathrm{norm.}})$ but also to a change of the
overall form of this dependence. The additional peaks arising on the $I$-$V$
curves can be attributed to the ferromagnetic resonance and the nonzero
coupling between Josephson and magnetic moment oscillations.
In order to observe these peaks experimentally, one should perform
measurements with different samples that contain ferromagnetic layers of
varying thickness. Then, according to our theoretical result, one would be
able to differentiate between ordinary Fiske steps and peaks caused by
interaction of Josephson and magnetic oscillations in the F layers.
Note that in the limit of a very short junction ($L/l_J \ll 1$) there is no
coupling between Josephson and magnetic moment oscillations. Indeed, in this
limit we obtain from Eq.~(\ref{eq-ResultForEta})

\begin{equation}
\delta\eta = \mathrm{Im} \left\{ \frac{\Omega_J^2}{\Omega_V\left(
\Omega_V-i\gamma_R\right)} \right\}.  \label{eq-EtaLimitStoZero}
\end{equation}
It is seen that magnetic characteristics such as $\Omega_{M}$ of the F layers
drop out from this expression.

In the limit of long junctions, $L/l_J \gg 1$, the expression for the
current correction can be approximated by

\begin{eqnarray}
\delta\eta
&=&
\mathrm{Im} \left\{ \frac{1}{P_{\Omega}(V,H)} \right\}
\label{eq-EtaLimitLtoInfty}
\\
&=&
\mathrm{Im} \left\{ \frac{\Omega_V^2-i\gamma_R \Omega_V}{\Omega_J^2} -
\kappa_H^2l_J^2 \left[ 1 + \frac{s}{(1+s)\mathcal{L}_{\Omega_VF}}\right]
\right\}^{-1}.
\nonumber
\end{eqnarray}
In accordance to Fig.~\ref{fig-LongJJcurrentCorrection2} we obtain for $s = 0$
a single peak at normalized voltage $V_{\mathrm{norm.}} = \kappa_H l_J$ while for
$s\neq0$ and $\kappa_Hl_J = \Omega_M/\Omega_J$ there exist two peaks at

\begin{equation}
V_{\mathrm{norm.}} = \frac{\Omega_M}{\Omega_J} \sqrt{(1+s) \pm \sqrt{s(1+s)}}.
\end{equation}
Finally, for the general case $s\neq0$ and $\kappa_Hl_J \neq \Omega_M/\Omega_J$ we find
in leading order in the parameter $s$ two peaks located at normalized voltages

\begin{subequations}
\begin{eqnarray}
V_{\mathrm{norm.}}^{(1)}
&=&
\kappa_H l_J \sqrt{1 + s \cdot \frac{x^2}{1-x^2}}
\\
V_{\mathrm{norm.}}^{(2)}
&=&
\frac{\Omega_M}{\Omega_J} \sqrt{1 + s \cdot \frac{1-2x^2}{1-x^2}}
\end{eqnarray}
\end{subequations}
where $x = (\Omega_M/\Omega_J)/(\kappa_H l_J)$. These analytical expressions perfectly describe
the peak locations of the current-voltage characteristics in the limit $L/l_J \gg 1$ as exemplarily
shown in Fig.~\ref{fig-LongJJcurrentCorrection2} for junctions with $L/l_J = 10$.

\section{Coupled Collective Modes}

\label{section-CoupledModes}

In this section, we analyze the spectrum of coupled collective modes in long
Josephson junctions with a ferromagnetic layer. So far we have derived essentially two
(coupled) equations, Eqs.~(\ref{eq-VarphiDynamicsAndM}) and (\ref%
{eq-LLGvarhiCoupled}), that describe respectively the dynamics of the phase
difference $\varphi$ of the S layers and the magnetization $\mathbf{M}$
of the ferromagnetic layers. Here, we consider again the case when the
magnetization $\mathbf{M}_0$ is aligned normal to the interface so that in
equilibrium $\mathbf{B}_0 = 0$. Small perturbations near the equilibrium
result in precessional motion of the magnetic moment $\mathbf{M}$ and in a
variation of the phase difference $\varphi$ in space and time. In order to
find the spectrum of collective modes in the system, we represent the phase
difference $\varphi$ and the magnetic moment $\mathbf{M}$ in the form

\begin{equation}
\varphi = \varphi_0 + \psi, \quad \mathbf{M} = M_0\mathbf{n}_z + \mathbf{m}%
_{\perp},
\end{equation}
where $\mathbf{n}_z$ is the unit vector normal to the SF interface and the
functions $\psi$ and $\mathbf{m}_{\perp}$ are assumed to be small, $|\psi|
\ll |\varphi_0|$ and $|\mathbf{m}_{\perp}| \ll |M_0|$. Linearizing Eq.~(\ref%
{eq-VarphiDynamicsAndM}) with respect to $\psi$, we find that the function $%
\psi(x,t)$ obeys the equation

\begin{eqnarray}
&& \Omega_J^{-2} \left( \frac{\partial^2 \psi}{\partial t^2} + \gamma_R
\frac{\partial \psi}{\partial t} \right) - l_J^2 \nabla_{\perp}^2\psi + \psi
=  \label{eq-CollModes-ODEVarphi} \\
&& \qquad\qquad\qquad \qquad\qquad\qquad = \frac{c\widetilde{d}_F}{2%
\widetilde{\lambda}_L j_c} \Big[\nabla \times \mathbf{m}_{\perp}\Big]_z.
\notag
\end{eqnarray}
The perturbation $\mathbf{m}_{\perp}$ of the magnetic moment is parallel to
the SF interface and is described by the equation

\begin{eqnarray}
\frac{\partial \mathbf{m}_{\perp}}{\partial t} &=& \Omega_M \Bigg\{ \left(
1+s-l_M^2 \nabla_{\perp}^2 \right) \Big[ \mathbf{n}_z \times \mathbf{m}%
_{\perp} \Big] -  \label{eq-CollModes-ODEm} \\
&& \phantom{\Omega_M \Bigg\{} - \frac{\Phi_0}{(4\pi)^2 \beta \widetilde{%
\lambda}_L} \nabla_{\perp} \psi \Bigg\} + \gamma_M \left[ \mathbf{n}_z
\times \frac{\partial \mathbf{m}_{\perp}}{\partial t} \right],  \notag
\end{eqnarray}
where we included again the Gilbert damping term, which was neglected in Eq.~(%
\ref{eq-LLGvarhiCoupled}). Fourier transforming the perturbations $\varphi(%
\mathbf{r},t)$ and $\mathbf{m}_{\perp}(\mathbf{r},t)$ to $(\mathbf{k}%
_{\perp},\omega)$ representation and combining Eqs.~(\ref%
{eq-CollModes-ODEVarphi}) and (\ref{eq-CollModes-ODEm}) into a single
equation, we obtain

\begin{gather}
\mathcal{M} \left(
\begin{array}{c}
\varphi(\mathbf{k}_{\perp},\omega) \\
\mathbf{m}_{\perp}(\mathbf{k}_{\perp},\omega)%
\end{array}
\right) = 0,  \label{eq-CollModes-MatrixM} \\
\mathcal{M} =
\begin{pmatrix}
\Omega_J^{-2}\left( \omega_J^2 - \omega^2 \right) & i b k_y & - i b k_x \\
i a k_x & - i \omega & \omega_M \\
i a k_y & - \omega_M & - i \omega%
\end{pmatrix},
\end{gather}
where $\omega_J^2 \equiv \omega_J^2(k,\omega) = \Omega_J^2\left(1+k^2
l_J^2\right) - i\gamma_R\omega,\, \omega_M \equiv \omega_M(k,\omega) =
\Omega_M \left( 1+s+k^2 l_M^2 \right) - i\gamma_M\omega$, $b = s\beta c /j_c$,
$a = s \Omega_M l_J^2 /b$, and $k=|\mathbf{k}_{\perp}|$.

The homogeneous equation~(\ref{eq-CollModes-MatrixM}) has a non-vanishing
solution provided the determinant of $\mathcal{M}$ equals zero. Setting $%
\mathrm{det}(\mathcal{M})$ equal to zero we obtain the dispersion relation

\begin{equation}
\left[ \omega^2 - \omega_J^2 \right] \left[ \omega^2 - \omega_M^2 \right] =
sv_J^2\Omega_M\omega_M k^2.  \label{eq-CollModes-DispRel}
\end{equation}

\begin{figure}[t!]
\includegraphics[scale=0.25]{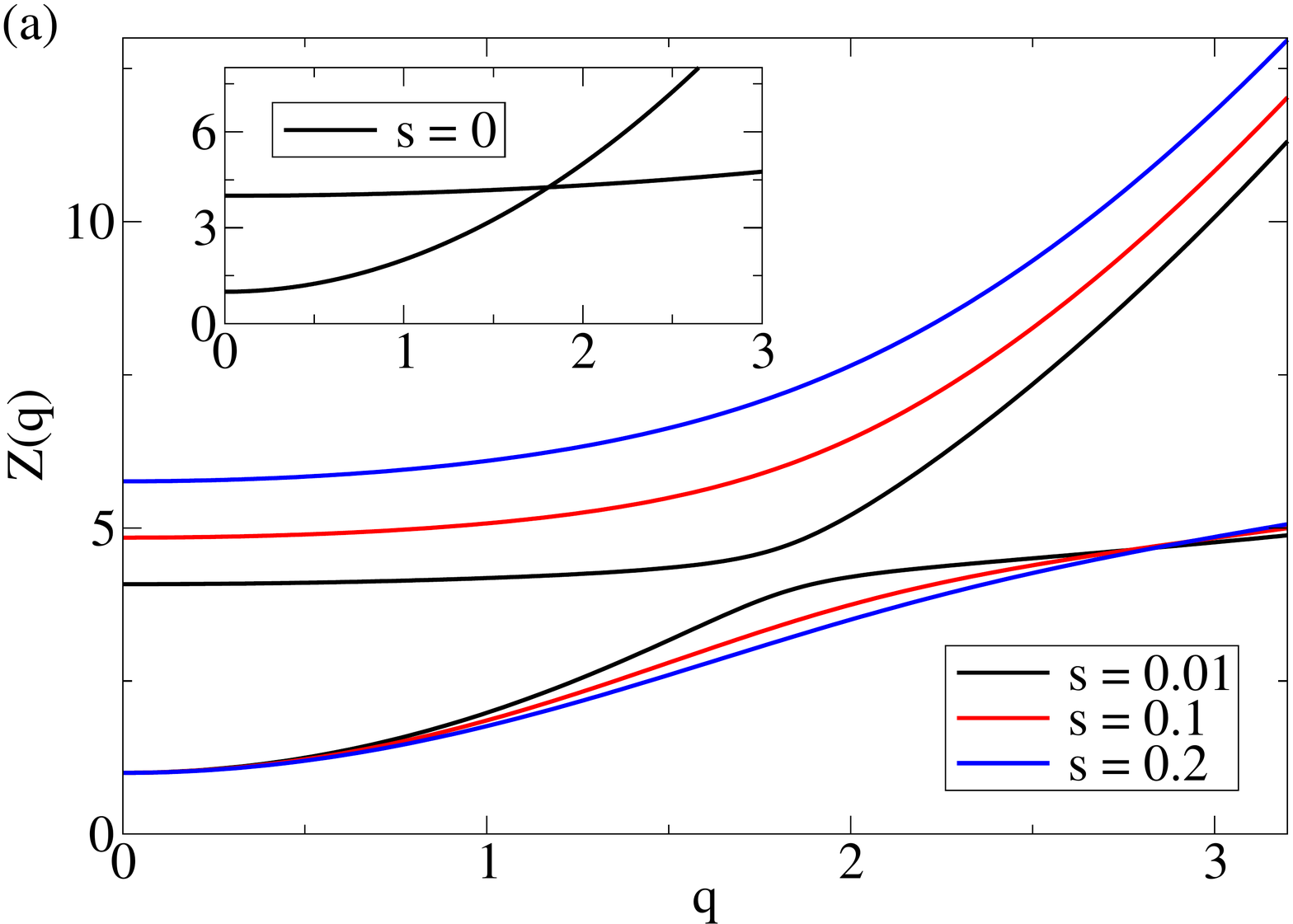}\newline
\includegraphics[scale=0.25]{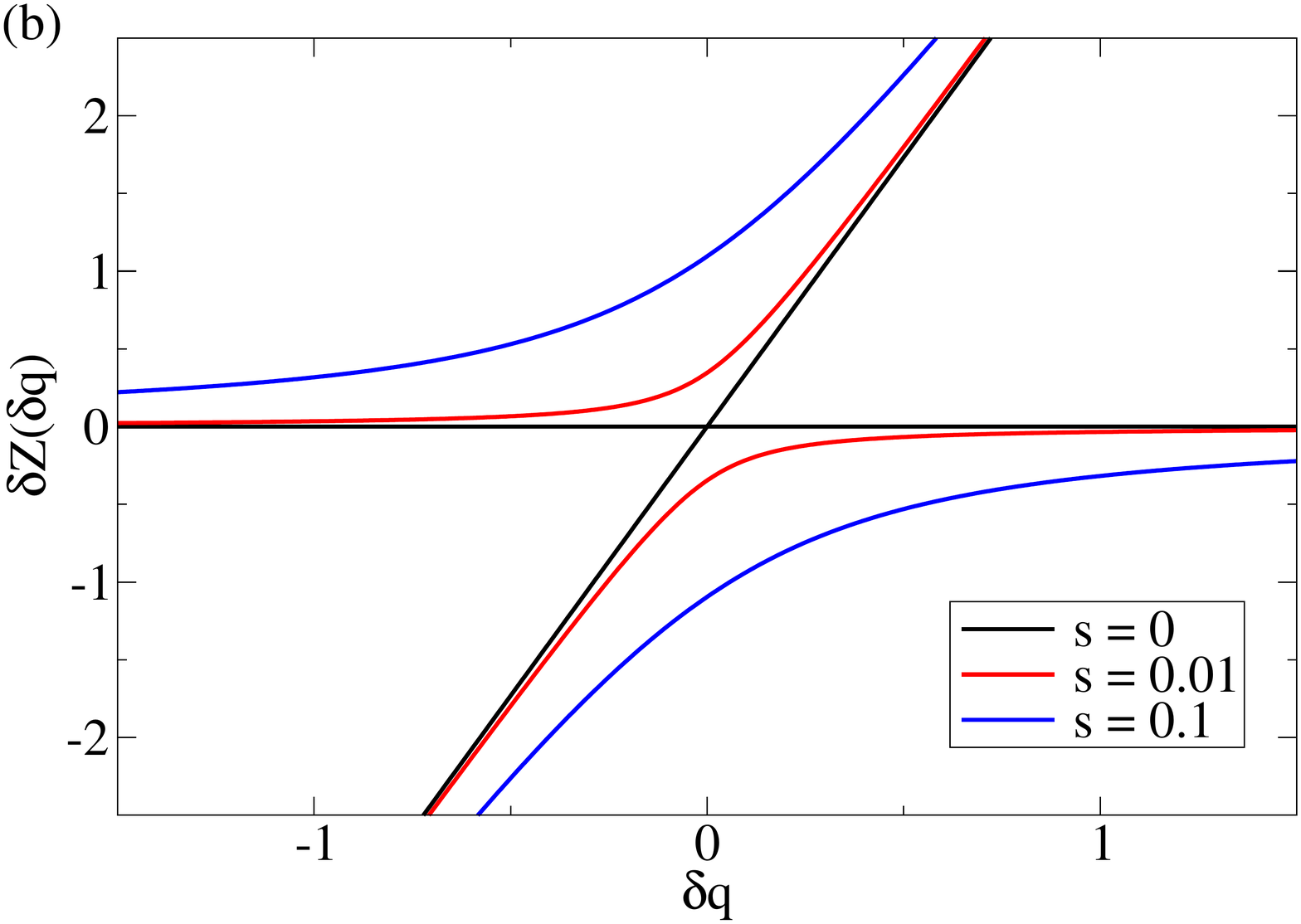}
\caption{(Color online) Spectrum of coupled spin and plasma-like modes. (a)
The function $Z(q)=(\protect\omega(q)/\Omega_J)^2$ determined by Eq.~(\ref%
{eq-CollModes-DispRelDimless2}) is shown for finite values of $s$ and in the
inset for the case $s=0$. (b) The dependence $\protect\delta Z(\protect%
\delta q)$ that represents the spectrum close to the crossing point is
plotted according to Eq.~(\ref{eq-CollModes-DispRelCrossPoint}). The
following parameters are chosen: $Z_M=(\Omega_M/\Omega_J)^2=4,\, l_M/l_J=0.1$%
.}
\label{fig-CoupledModesSpectrum}
\end{figure}

From Eq.~(\ref{eq-CollModes-DispRel}) we can conclude that the spin and
charge excitations decouple only in the limit when the right-hand side of
this equation can be neglected. In this case the spin waves with
spectrum $\omega_M(k, \omega)$ and the plasmalike Josephson waves with
spectrum $\omega_J(k, \omega)$ exist separately. In the general case, Eq.~(%
\ref{eq-CollModes-DispRel}) describes the spectrum of coupled spin waves and
plasma-like modes in the system. The most interesting behavior corresponds
to the case $\Omega_M > \Omega_J$. In this situation, the two branches of the
spectrum cross each other in the absence of the coupling, while a finite
coupling leads to mutual repulsion of these branches.

In order to show this explicitly, we consider the case without damping, $%
\gamma_R = \gamma_M = 0$, and assume that $s \ll 1$ and $l_M \ll l_J$,
which means that we neglect the spatial dispersion of spin waves on the
Josephson length (these conditions are usually fulfilled experimentally).
It is convenient to write Eq.~(\ref{eq-CollModes-DispRel}) in the
dimensionless form

\begin{equation}
\left[ Z-1-q^2 \right] \left[ Z-Z_M \right] = sq^2 Z_M,
\label{eq-CollModes-DispRelDimless}
\end{equation}
where $Z = (\omega/\Omega_J)^2, \, Z_M = (\Omega_M/\Omega_J)^2$ and $q = l_J
k$. One can see that for $s = 0$ the two dispersion curves $Z = 1+q^2$ and $%
Z = Z_M$ cross each other at $q_0^2 = Z_M-1$. To find the form of the dispersion
curve in the vicinity of the crossing point $q_0$, we represent $Z$ and $q$,
respectively, as $Z = Z_M + \delta Z$ and $q = q_0 + \delta q$. Then, one can
easily obtain from Eq.~(\ref{eq-CollModes-DispRelDimless})

\begin{equation}
\delta Z = q_0 \left[ \delta q \pm \sqrt{\delta q^2 + sZ_M} \right].
\label{eq-CollModes-DispRelCrossPoint}
\end{equation}
In Fig.~\ref{fig-CoupledModesSpectrum} we plot the spectrum of coupled
spin and plasma-like modes and the function $\delta Z(\delta q)$ close to
the crossing point. Here, we take into account a finite value of the
parameter $l_M/l_J$ so that Eq.~(\ref{eq-CollModes-DispRelDimless}) that
determines the function $Z(q)$ takes the form

\begin{equation}
\left[ Z-1-q^2 \right] \left[ Z-\widetilde{Z}_M \right] = sq^2 \sqrt{%
\widetilde{Z}_M} Z_M,  \label{eq-CollModes-DispRelDimless2}
\end{equation}
with $\widetilde{Z}_M = Z_M [1+s+q^2(l_M/l_J)^2]^2$. The inset of Fig.~\ref%
{fig-CoupledModesSpectrum}(a) indicates that the two branches indeed cross
each other for $s=0$, whereas for $s\neq 0$ we find a \textquotedblleft
repulsion\textquotedblright of the spin and Josephson excitations. Figure~\ref%
{fig-CoupledModesSpectrum}(b) displays the function $\delta Z(\delta q)$
that represents the behavior of the spectrum in the vicinity of the crossing
point and distinctly emphasizes the mutual repulsion. Both the dispersion
curves $Z(q)$ and $\delta Z(\delta q)$ given by Eqs.~(\ref%
{eq-CollModes-DispRelCrossPoint}) and (\ref{eq-CollModes-DispRelDimless2}),
respectively, are presented for several values of $s$ and the parameters $%
\Omega_M/\Omega_J = 2, \, l_M/l_J = 0.1$.

\section{Ferromagnetic Resonance}

\label{section-FR}

\begin{figure*}[t]
\includegraphics[scale=0.25]{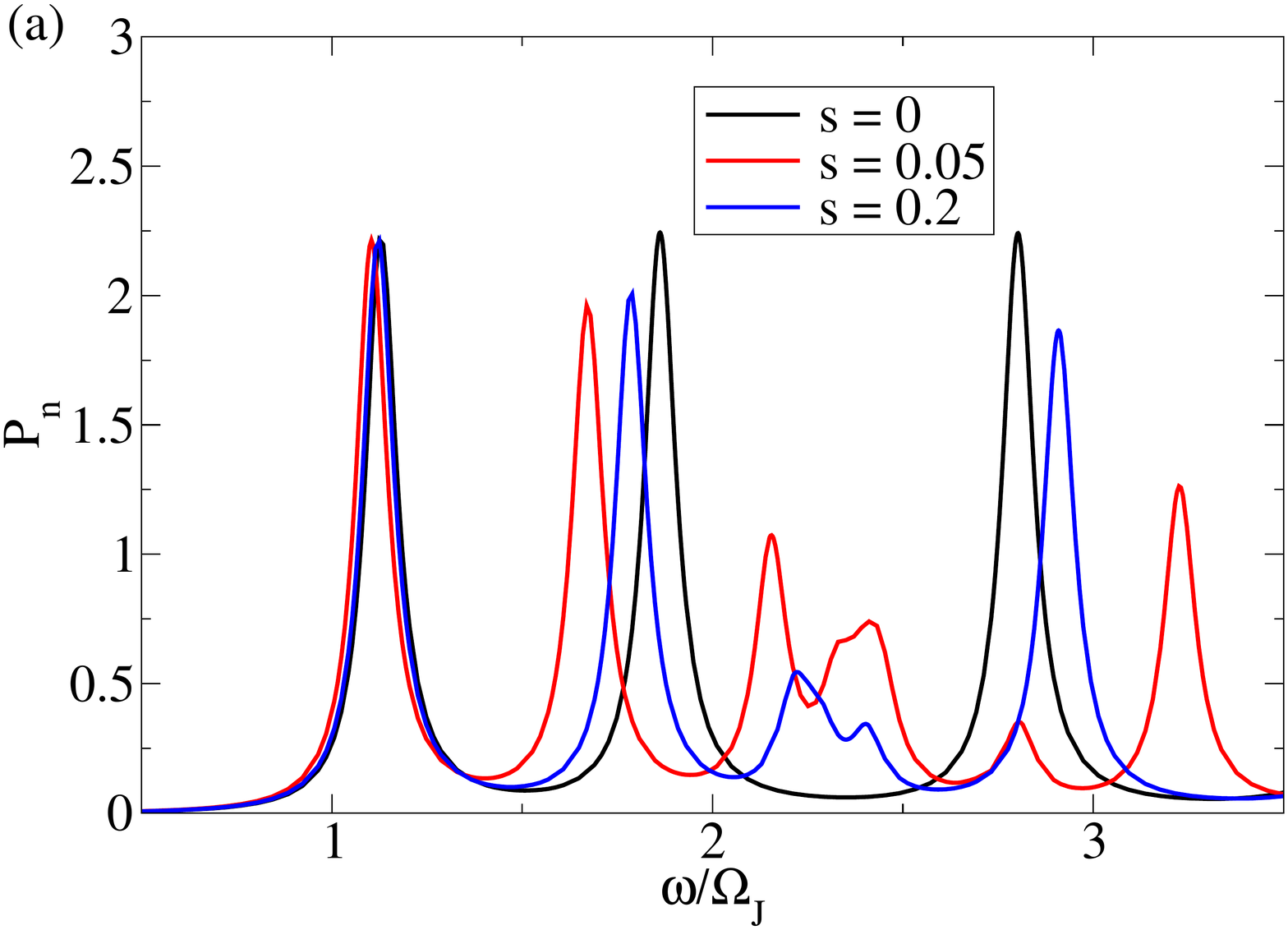}\qquad %
\includegraphics[scale=0.25]{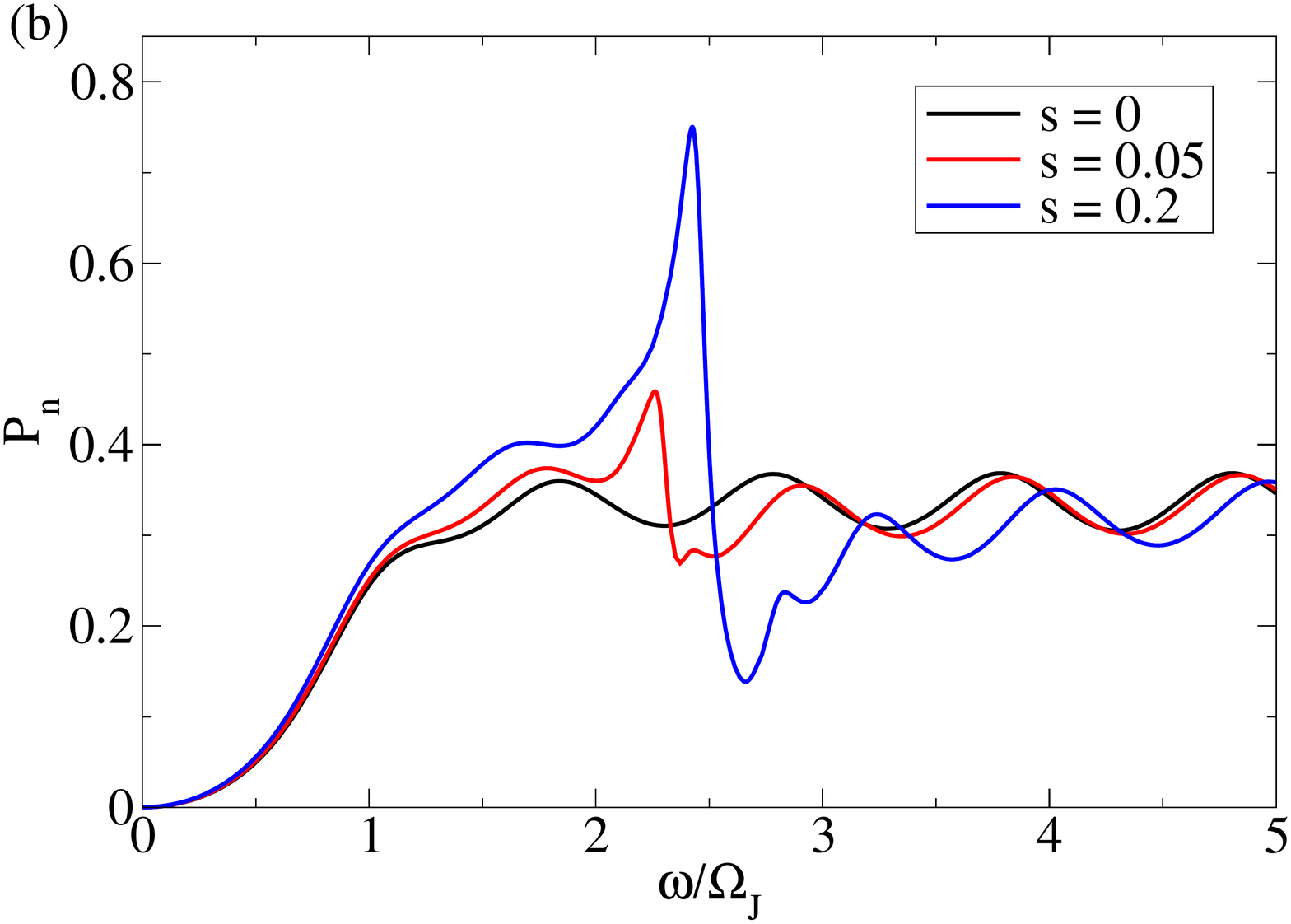} %
\includegraphics[scale=0.25]{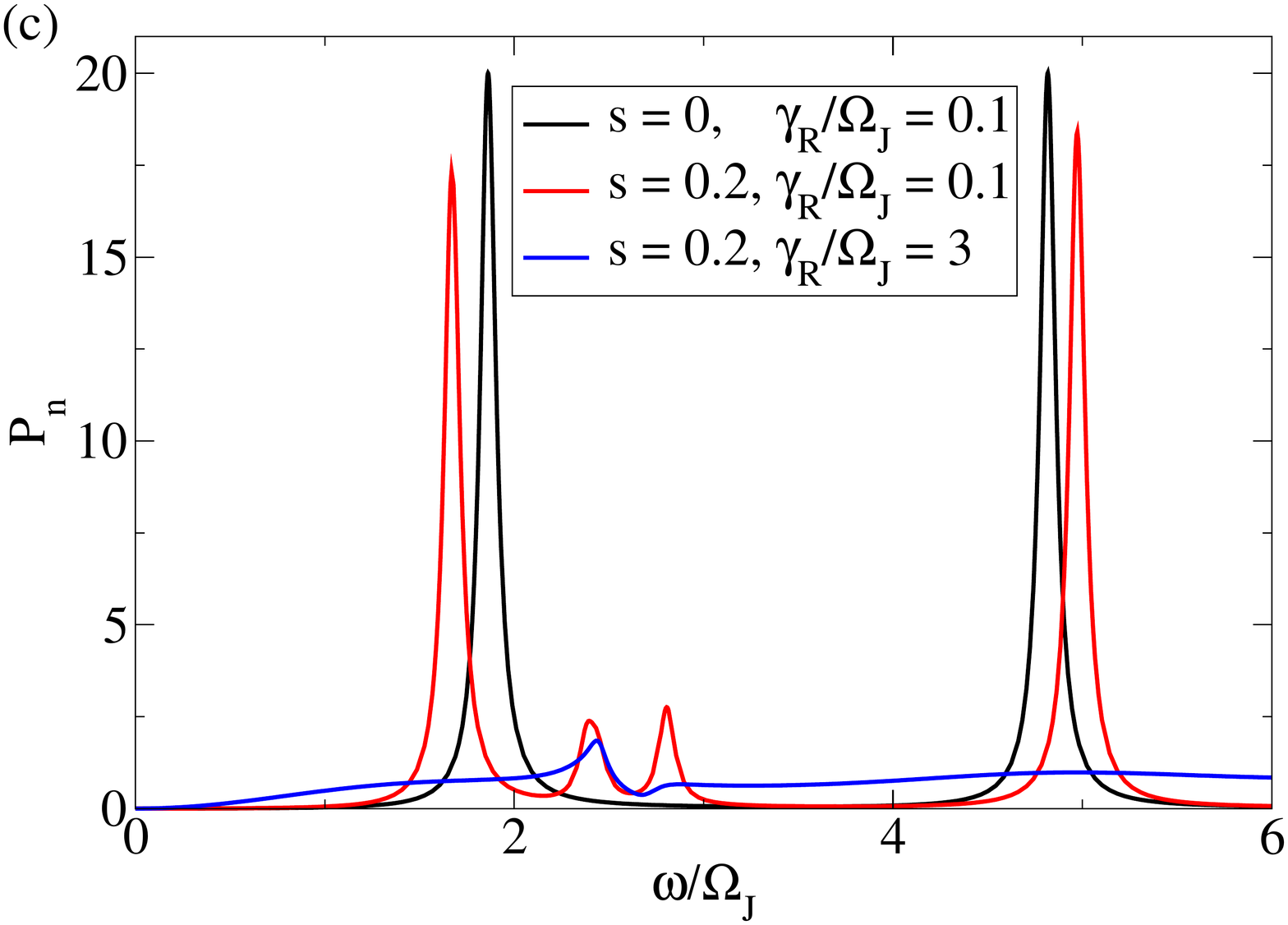}\qquad %
\includegraphics[scale=0.25]{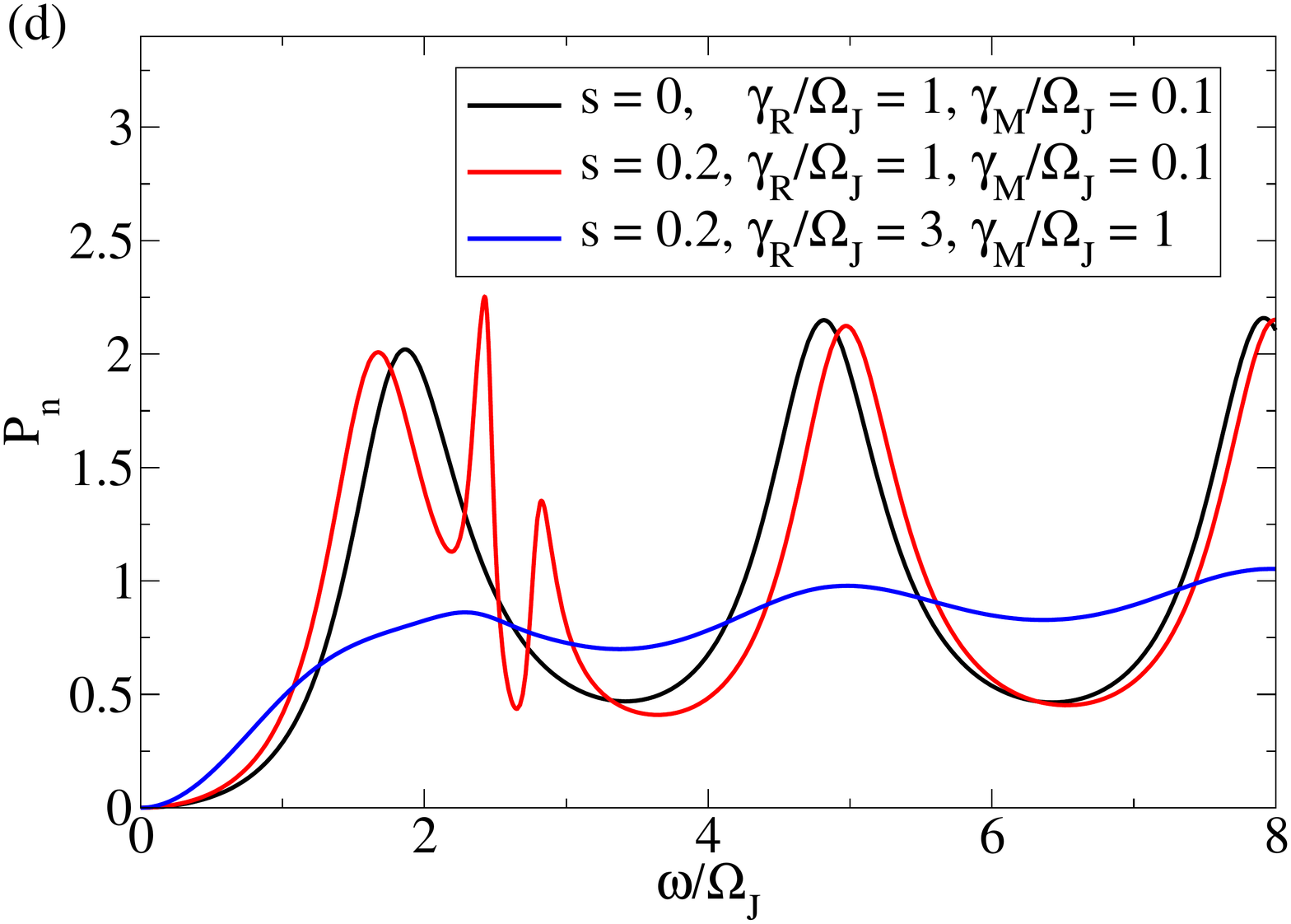}
\caption{(Color online) Frequency dependence of the normalized absorbed
power $\mathcal{P}_n$ in the JJ as a function of the normalized frequency $%
\protect\omega/\Omega_J$ for different values of the parameter $s$ and for
different damping coefficients $\gamma_R/\Omega_J$ and $\gamma_M/\Omega_J$.
The figures (a) and (b) are shown for the parameters $L_n=3,
\protect\gamma_M/\Omega_J=0.1, (\Omega_M/\Omega_J)^2=5$ and, respectively, (a)
$\protect\gamma_R/\Omega_J=0.1$, (b) $\protect\gamma_R/\Omega_J=1$. With
regard to figures (c) and (d) we have chosen $L_n=1, (\Omega_M/\Omega_J)^2=5$
and in (c) $\protect\gamma_M/\Omega_J=0.1$.}
\label{fig-FR_Pn_1}
\end{figure*}

In this section, we study the response of the system to an external
oscillating magnetic field $H_{\mathrm{ext}}(t) = H_{\nu} \sin(\nu t)$ with
a small amplitude $H_{\nu} \ll M_0$ and frequency $\nu$. The applied field
is supposed to be directed along the $y$ axis, i.e., $\mathbf{H}_{\mathrm{ext}%
}(t) = H_{\mathrm{ext}}(t) \mathbf{n}_y$. We assume again that the
equilibrium magnetization $\mathbf{M}_0$ is oriented in the $z$ direction, $%
\mathbf{M} = M_0 \mathbf{n}_z$. The external magnetic field $H_{\mathrm{ext}%
}(t)$ causes precessional motion of the magnetization vector $\mathbf{M}$
and a variation of the phase difference $\varphi$ in space and time. As
before (see Sec.~\ref{section-LongFiske}), we, respectively, represent
magnetization and phase difference in the form $\mathbf{M}(x,t) = M_0
\mathbf{n}_z + m_y(x,t) \mathbf{n}_y$ and $\varphi(x,t) = \varphi_0 +
\psi(x,t)$. Here, $\varphi_0$ is a constant determined by a bias current $%
j_b = j_c \sin(\varphi_0)$ and $\psi(x,t), \,m_y(x,t)$ are small
perturbations due to the external ac magnetic field $H_{\mathrm{ext}}(t) =
H_{\nu}\, \mathrm{Im}\left[ \exp(i\nu t) \right]$, $|\psi| \ll
|\varphi_0|,\, |m_y| \ll |M_0|$. Due to the coupling of $\mathbf{M}$ and $%
\varphi$, we expect modifications of the ferromagnetic resonance in the
system appearing as additional features in absorption spectra.

Thus, to study ferromagnetic resonance, we need to calculate the power $%
\mathcal{P}$ (per unit area) absorbed in the system. The absorbed power $%
\mathcal{P}$ can be found as the time-averaged difference between the energy
flux $\mathcal{S}_{\mathrm{in,out}}$ coming in and out of the system. These
fluxes are expressed in terms of Poynting vectors $\mathcal{S}$\cite{LL_ED}

\begin{equation}
\mathcal{P} = \int dzdy\, \mathbf{n}_x \cdot \left\langle \mathcal{S}_{%
\mathrm{in}} - \mathcal{S}_{\mathrm{out}} \right\rangle,
\label{eq-FRenergyFlux}
\end{equation}
where $\mathcal{S}_{\mathrm{in,out}} = (c/4\pi) [\mathbf{E} \times \mathbf{H}%
]_{x = \pm L}$ and the angular brackets denote averaging with respect to
time $t$.

The electric field $\mathbf{E}=\mathbf{n}_{z}E$ is directed along the $z-$%
axis and is related to the time derivative of the phase difference via the
Josephson relation

\begin{equation}
E=-(1/d)(\hbar /2e)\partial \psi /\partial t.
\label{eq-ElectricField}
\end{equation}
Therefore, in order to find the Poynting vector $\mathcal{S}$,
we have to calculate the function $\psi(t)$ which is determined
by an applied weak ac magnetic field $H_{\mathrm{ext}}(t)$. This
vector $\mathcal{S}$ differs from zero only in the insulating
layer of thickness $d$. The magnetic field consists only of the
applied ac field, $\mathbf{H}(x=\pm L)=H_{\mathrm{ext}}(t)\mathbf{n}_{y}$
and, therefore, the Poynting vectors are directed parallel to the $x-$axis.
We represent the phase difference in form of the Fourier
transform $\psi (x,t)=\int d\omega /(2\pi )\exp (i\omega t)\psi (x,\omega),\,
\psi (x,-\omega )=\psi ^{\ast }(x,\omega )$.

The function $\psi(x, \omega)$ obeys an equation that is derived in a way
similar to the derivation of Eqs.~(\ref{eq-FinalODEvarphi})--(\ref{eq-CurrentEtaPsiAvg})
and has the form

\begin{equation}
\frac{\partial^2}{\partial x_n^2} \psi(x_n, \omega) - \kappa_{\omega}^2
\psi(x_n,\omega) = 0,  \label{eq-FRodePsi}
\end{equation}
where we have introduced the dimensionless variable $x_n = x/l_J$ and have
set

\begin{equation}
\kappa_{\omega}^2 = \frac{\mathcal{L}_{\omega J}}{1+s/[(1+s)\mathcal{L}%
_{\omega F}]} \equiv \frac{\mathcal{L}_{\omega J}}{a_{\omega}}
\label{eq-FRdefKappa}
\end{equation}
with $\mathcal{L}_{\omega J}~\equiv~\cos(\varphi_0)~-~\omega(\omega-i\gamma_R)/\Omega_J^2$,\,
$a_{\omega}~=~1+s/[(1+s)\mathcal{L}_{\omega F}]$, and $\mathcal{L}_{\omega F}$ is defined in
Eq.~(\ref{eq-DefLomegaF}).
\begin{figure*}[t!]
\includegraphics[scale=0.25]{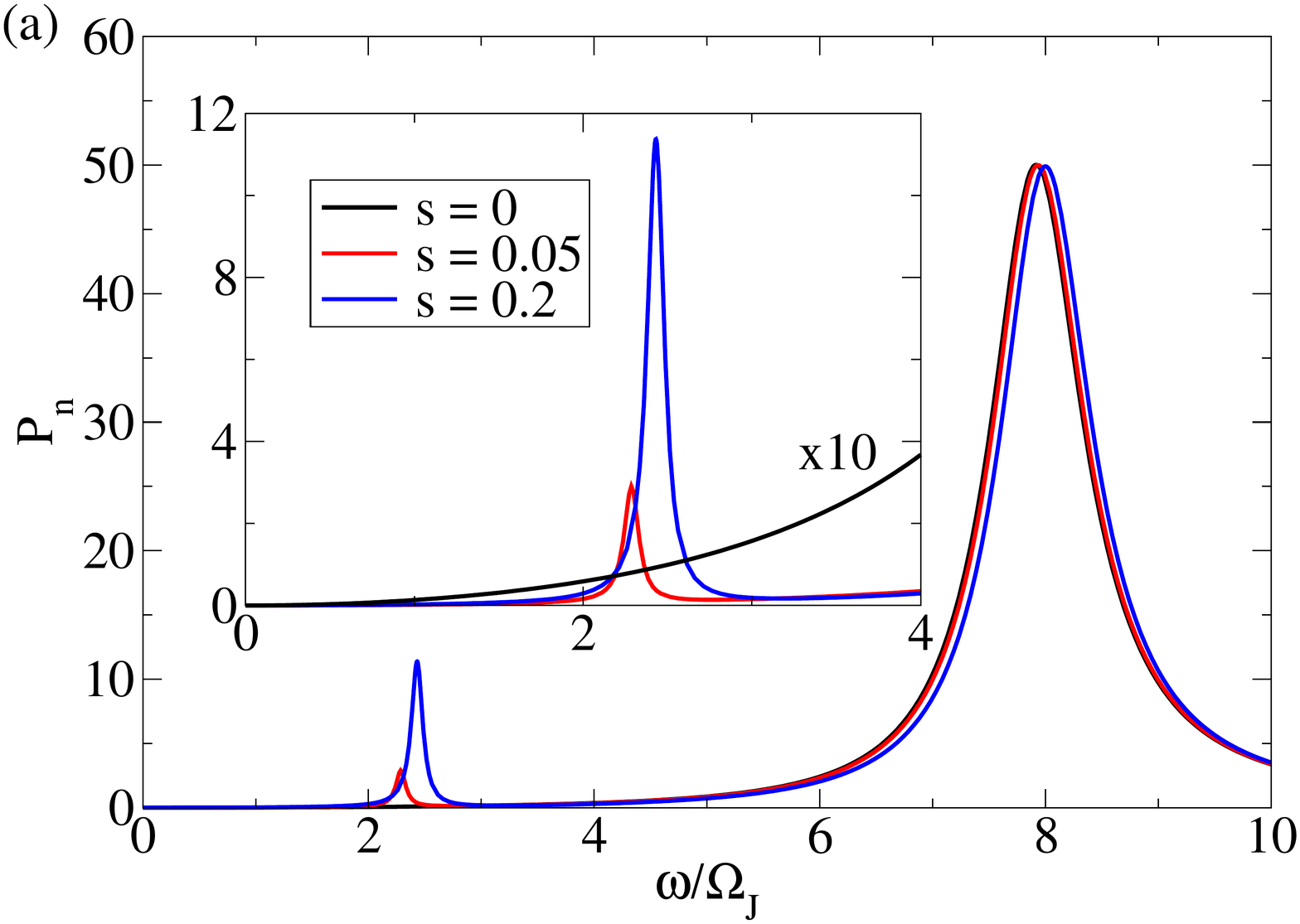} \qquad %
\includegraphics[scale=0.25]{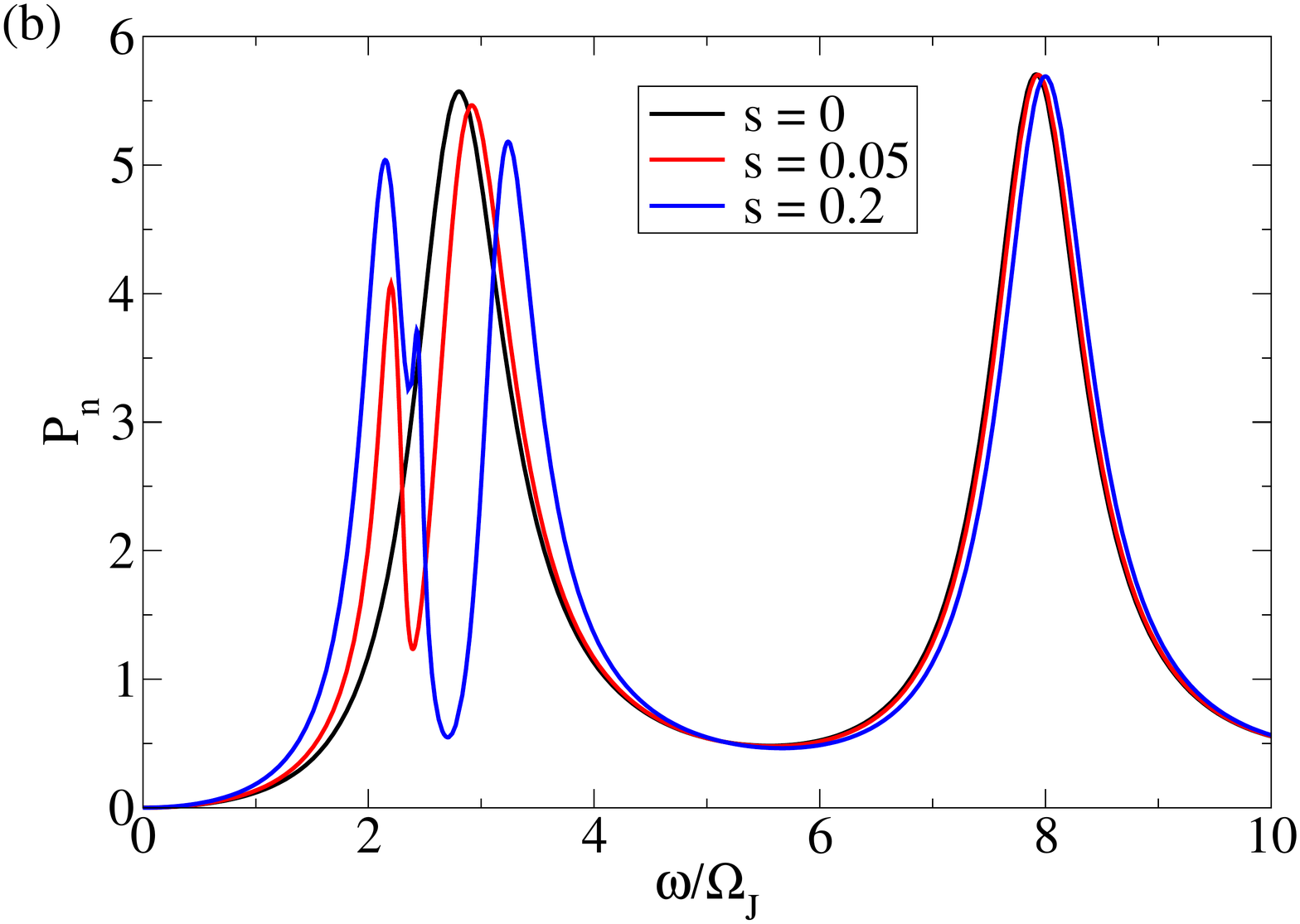}
\caption{(Color online) Frequency dependence of the normalized absorbed
power $\mathcal{P}_n$ as a function of the normalized frequency $\protect%
\omega/\Omega_J$ for really short junctions. Here, too, the figures are
displayed for different values of the parameter $s$ and for the choice $%
\protect\gamma_R/\Omega_J=1, \protect\gamma_M/\Omega_J=0.1,
(\Omega_M/\Omega_J)^2=5$. In figure (a) $L_n=0.2$ whereas in figure (b) $%
L_n=0.6$.}
\label{fig-FR_Pn_2}
\end{figure*}
Equation~(\ref{eq-FRdefKappa}) is supplemented by the boundary conditions

\begin{equation}
\left. a_{\omega} \frac{\partial}{\partial x_n} \psi(x, \omega)
\right|_{x=\pm L} = -\frac{H_{\mathrm{ext}}(\omega)}{H_0}\, \mathrm{and} \,
H_0 = \frac{\Phi_0}{4\pi \widetilde{\lambda}_L l_J}
\label{eq-FRbcPsi}
\end{equation}
that can be obtained from Eqs.~(\ref{eq-BinSlayer}) and (\ref%
{eq-BinSlayerPrefactor}). As a consequence, the solution for Eq.~(\ref%
{eq-FRodePsi}) has the form

\begin{equation}
\psi(\pm L, \omega) = \mp \frac{H_{\mathrm{ext}}(\omega) L_n}{H_0} \frac{%
\tanh(\theta_{\omega})}{a_{\omega} \theta_{\omega}}  \label{FR7}
\end{equation}
where $\theta_{\omega} = \kappa_{\omega} L_n \equiv \theta_{\omega}^{\prime}
+ i\theta_{\omega}^{\prime\prime},\, L_n = L/l_J$. Fourier transforming Eq.~(%
\ref{FR7}) back into the time representation we obtain

\begin{equation}
\psi(\pm L,t) = \mp\frac{H_{\nu} L_n}{H_0} \mathrm{Re}\left[ \frac{e^{i\nu t}%
} {a_{\nu} \theta_{\nu}} \tanh(\theta_{\nu}) \right].  \label{FRa8}
\end{equation}
Taking into account that all the quantities do not depend on $y$, the
absorbed power $\mathcal{P}$ can be represented as

\begin{equation}
\mathcal{P} = 2L_y \frac{\hbar c}{4\pi e}\left\langle \frac{\partial
\psi(L,t)}{\partial t} H_{\mathrm{ext}}(t) \right\rangle,
\label{FRb8}
\end{equation}
where $L_y$ is the length of the junction along the $y$ direction.
Substituting Eq.~(\ref{FRa8}) into Eq.~(\ref{FRb8}) and relabeling the
external field frequency $\nu \rightarrow \omega$, we finally arrive at

\begin{equation}
\mathcal{P} = \frac{\Phi_0 H_0}{(2\pi)^2} \left(\frac{H_{\omega}}{H_0}%
\right)^2 \frac{L_x L_y}{l_J}\, \omega \,\, \mathrm{Im}\left[ \frac{%
\tanh\left(\theta_{\omega}\right)}{a_{\omega} \theta_{\omega}} \right],
\label{FRa9}
\end{equation}
where $L_{x} \equiv L$. This formula differs drastically from a
standard formula for the absorbed power $\mathcal{P}$ in ferromagnetic
films because it describes the power absorption not only in the F film, but
also in the Josephson junction. In particular, $\mathcal{P}\neq 0$ even in the
absence of the ferromagnetic layer. In this case, Eq.~(\ref{FRa9}) describes
the power needed to excite standing plasma waves.

In Fig.~\ref{fig-FR_Pn_1}, we plot the frequency dependence of the normalized
absorbed power $\mathcal{P}_n = (\omega/\Omega_J) \mathrm{Im}\left[%
\tanh(\theta_{\omega})/(a_{\omega} \theta_{\omega}) \right]$ as a function
of normalized frequency $\omega/\Omega_J$ at different $s$ and
normalized junction length $L/l_J$.
Generally speaking, from Fig.~\ref{fig-FR_Pn_1}(a), we see that at $s=0$ (no
ferromagnetic layer), there are periodic resonances related to excitation of
standing waves in the Josephson junction (Josephson plasma resonances).
Interestingly, in the presence of the ferromagnetic layers, $s\neq 0$,
additional peaks appear on the curves. These peaks are caused by the
ferromagnetic resonance in the F layer at frequencies $\omega \approx
\Omega_M$. With increasing $s$ the influence of the F layer becomes more and
more pronounced. One can see this from the fact that, for instance, the
spectrum close to $\omega \approx \Omega_M$ appears to have a more complicated
structure and the peaks increase in height.

To indicate the influence of the (normalized) damping parameters $\gamma_R$
and $\gamma_M$, we display in Fig.~\ref{fig-FR_Pn_1}(b) the normalized
absorbed power $P_n$ for the case $\gamma_R \gg \gamma_M$. This shows that
the periodic resonances in the junction are strongly suppressed and the
absorption spectrum is dominated by the effect of the ferromagnetic layer.
In addition to that, the normalized length of the junction $L_n = L_x/l_J$
also determines the absorption spectrum. In Figs.~\ref{fig-FR_Pn_1}(c) and
\ref{fig-FR_Pn_1}(d), $P_n$ is shown for the case $L_n~=~1$ and here, too, for different
values of the parameters $s, \gamma_R,$ and $\gamma_M$. We find that the
distance between periodic resonances is larger for short junctions and
compared to Fig.~\ref{fig-FR_Pn_1}(a), where $L_n~=~3$, the influence of the F
layer on the absorption spectrum is weaker. The blue curve in Fig.~\ref%
{fig-FR_Pn_1}(c) reveals that the periodic resonances can be almost
completely suppressed by increasing the damping parameter $\gamma_R$.
Eventually, Fig.~\ref{fig-FR_Pn_1}(d) indicates that in systems where both
damping parameters $\gamma_R$ and $\gamma_M$ are large and of the same order of
magnitude, the effect of the ferromagnetic layer becomes negligible.

In Fig.~\ref{fig-FR_Pn_2}, we show the frequency dependence of the absorbed
power $P_n$ for short junctions of length $L_n = 0.2$ [see Fig.~\ref%
{fig-FR_Pn_2}(a)] and $L_n = 0.6$ [see Fig.~\ref{fig-FR_Pn_2}(b)]. In Fig.~\ref%
{fig-FR_Pn_2}(a), the peak at $\omega/\Omega_J \approx 2.3$ is related to
the ferromagnetic resonance in the F film. In contrast to this, slightly
longer junctions feature a much stronger influence of the ferromagnetic
layer as becomes apparent from Fig.~\ref{fig-FR_Pn_2}(b). More importantly,
we find that the relative magnitudes of peaks due to the Josephson plasma
resonances and the ferromagnetic resonance are even in the case of $\gamma_R
\gg \gamma_M$ of the same order of magnitude for short junctions
contrary to Fig.~\ref{fig-FR_Pn_1}(b), $L_n = 3$, where the Josephson plasma
resonances are considerably smaller for the same choice of parameters.

Note that our analysis is valid for not too high frequencies as it is
assumed that the penetration depth is not frequency dependent. This means that
the inequality $\omega \ll \Omega_J l_J/\lambda_L = v_J / \lambda_L$ should
be fulfilled. For this reason, Figs.~\ref{fig-FR_Pn_2}(a) and \ref{fig-FR_Pn_2}(b) indicate
only a small number of Josephson plasma resonances.

\section{Discussion}
\label{section-Discussion}

We studied dynamic properties of Josephson junctions with a magnetically
active layer characterized by the magnetic susceptibility $\chi
(\omega ,k)$. These junctions may be of the SFIFS or SIFS type with
conducting or insulating ferromagnets. In the former case, we assumed that
both vectors $\mathbf{M}_{1}$ and $\mathbf{M}_{2}$ characterizing the
stationary orientation of magnetization in the F layers were aligned along
the $z$ direction, and our results are applicable only in this situation.

We calculated the form of the CVC for SFIFS junctions in the presence of a
weak magnetic field and found a modification of Fiske steps due to the
presence of the ferromagnetic layer. The position of these steps depends
on the relation between different parameters, especially between
$\kappa_{H}l_{J}$ and $\Omega_{M}/\Omega_{J}$.

We have also analyzed the spectrum of the collective coupled modes in long
JJs with a ferromagnetic layer. If the frequency of the ferromagnetic
resonance $\Omega_M$ is higher than the characteristic Josephson frequency $%
\Omega_J$, then coupled magneto-plasma modes (spin waves and Josephson
plasma-like modes) occur in the region of crossing terms.

The analysis of the ferromagnetic resonance in the F layer incorporated in
JJs of the SFS or SFIFS types shows that the peaks in the frequency
dependence of the absorbed power $P(\omega )$ correspond both to the
ferromagnetic resonance in the F film and to the Josephson plasma resonances
in the tunnel JJ.

It is not easy to compare our results with available experimental data. The
dynamic properties of ferromagnetic layers play a crucial role in
determining the form of the CVC (Fiske steps). Meanwhile, little is known
about these properties in experiments. It would be useful to study
experimentally magnetic resonance in the F layers at temperatures above the critical
temperature of the superconducting transition $T_{c}$. The frequencies of
the Josephson oscillations $\Omega _{J}$ and magnetic resonance $\Omega_{M}$
should not be very different. In addition, we assumed that the
easy-axis magnetization is perpendicular to the SF interface. There are no
data about magnetization orientation in junctions studied experimentally.

As to magnetic resonance, we are only aware of Refs.~\onlinecite{Aprili,ExpDyn1,ExpDyn2}
where ferromagnetic resonance was measured on SF structures. However, the
authors of Ref.~\onlinecite{Aprili} measured the CVC of a SFS junction with a
strong damping, but not the absorbed power. In Ref.~\onlinecite{ExpDyn1,ExpDyn2}, the
absorbed power was measured, however not in SIFS junctions, but in SF
bilayers. Thus further experiments are needed to study the interplay
between magnetic and Josephson oscillations in tunnel Josephson junctions
with a ferromagnetic layer.

\begin{center}
\textbf{ACKNOWLEDGMENT}
\end{center}

We thank SFB 491 for financial support.


\end{document}